\documentclass[manuscript]{aastex}
\usepackage{chngpage}
\usepackage{graphicx}
\usepackage{natbib}

\shorttitle{Spectral Evolution over Redshift of GRBs}
\shortauthors{Geng \& Huang}

\begin{document}

\title{On the Correlation of Low-energy Spectral Indices \\ and Redshifts of Gamma-ray Bursts}

\author{J. J. Geng\altaffilmark{1, 2} and Y. F. Huang\altaffilmark{1, 2}}

\altaffiltext{1}{Department of Astronomy, Nanjing University, Nanjing 210093, China; hyf@nju.edu.cn}
\altaffiltext{2}{Key Laboratory of Modern Astronomy and Astrophysics (Nanjing University), Ministry of Education, China}

\begin{abstract}
It was found by Amati et al. in 2002 that for a small sample of 9 gamma-ray bursts,
more distant events appear to be systematically harder in the soft gamma-ray band. Here, we
have collected a larger sample of 65 gamma-ray bursts, whose time integrated spectra
are well established and can be well fitted with the so called Band function. It is confirmed that
a correlation between the redshifts ($z$) and the low-energy indices ($\alpha$) of the Band function
does exist, though it is a bit more scattered than the result of Amati et al. This correlation can not be
simply attributed to the effect of photon reddening. Furthermore, correlations
between $\alpha$ and $E_{\rm peak}$ (the peak energy in the $\nu F_{\nu}$ spectrum in the rest frame), $\alpha$ and
$E_{\rm iso}$ (the isotropic energy release), $\alpha$ and $L_{\rm iso}$ (the isotropic luminosity) are
also found, which indicate that these parameters are somehow connected. The results may give
useful constraints on the physics of gamma-ray bursts.
\end{abstract}


\section{Introduction}
\label{sect:intro}

Correlation analysis plays an important role in leading us understanding the physics
of astronomical processes (Dyson \& Schaefer~\citeyear{dyson98}). The correlations between
the spectral parameters of the prompt emission of gamma-ray bursts
(GRBs) and other related parameters help to reveal their nature.
In this aspect, several empirical relationships have been found previously. The most
famous one is the correlation between the rest frame peak
energy ($E_{\rm peak}$, note that we will use $E_{\rm peak}^{\rm obs}$
to designate the corresponding peak energy in the observer's frame hereinafter)
of the $\nu F_{\nu}$ spectrum and the bolometric isotropic energy release ($E_{\rm iso}$) during the burst,
also known as the Amati relation (Amati et al.~\citeyear{amati02};
Amati~\citeyear{amati03}). Significant correlation is also found to exist between
$E_{\rm peak}$ and the collimation-corrected energy, which is usually called the
Ghirlanda relation (Ghirlanda et al.~\citeyear{ghir04}). Liang \& Zhang~\citeyearpar{liang05}
further established a three-parameter correlation between
$E_{\rm peak}$, $E_{\rm iso}$ and the jet-break time in the afterglow
light curve. The fourth relationship that should be mentioned is between $E_{\rm peak}$ and the
isotropic peak luminosity ($L_{\rm iso}$), i.e., the so called
Yonetoku relation (Yonetoku et al.~\citeyear{yone04}). Very
recently, for a sub-sample of long GRBs with known redshifts and with a
plateau phase in the afterglow, a tight three-parameter
correlation has also been found among the end time of the
plateau phase (in the GRB rest frame), the corresponding X-ray
luminosity and the isotropic $\gamma$-ray energy release (Xu \& Huang~\citeyear{XuHuang12}).

However, the nature of many of these relationships has been under debate since their
discovery. Some believe that they follow from the physics of the emission process during the
burst (Schaefer~\citeyear{schaefer04}; Bosnjak et al.~\citeyear{bosn08};
Amati et al.~\citeyear{amati09}; Ghirlanda et al.~\citeyear{ghir10}).
Especially, a few of the correlations (such as the Yonetoku relation) are found to
exist in time-resolved analysises of individual GRBs. The time-resolved behaviors are largely
similar for different GRBs and are also consistent with the time-integrated relations
(Ghirlanda et al.~\citeyear{ghir10}; Ghirlanda et al.~\citeyear{ghir11}; Nava et al.~\citeyear{nava12}).
An intrinsic physical origin for these correlations is thus strongly hinted, since it is quite unlikely
that instrumental selection effects or other observational biases can play the major role in a single burst.
However, it should be noted that observational selection effects still can not be completely excluded yet
(Nakar \& Piran~\citeyear{nakar05}; Butler et al.~\citeyear{butler07}; Shahmoradi \&
Nemiroff~\citeyear{shah11}).

The origin of the prompt emission of GRBs is still an open question. Within the standard fireball framework
(Piran~\citeyear{piran04}), many mechanisms (see Zhang \& M{\'e}sz{\'a}ros~2002 for a review) have been proposed
based on different ingredients of fireballs (e.g. kinetic energy dominated or Poynting flux dominated) and different
emission processes (e.g. synchrotron emission from internal shocks, photospheric emission etc.). It has been
suggested that both $E_{\rm peak}$ and $E_{\rm iso}$ should be somehow linked to the bulk Lorentz
factor ($\Gamma$) of the fireball in most scenarios (Amati 2006). This can potentially give an interpretation to
the Amati relation. Liang et al.~\citeyearpar{liang10} found a tight $\Gamma_{\rm0}$ -- $E_{\rm\gamma,iso}$
correlation by using the initial Lorentz factors constrained from afterglow analysises. They suggested
that the correlation may be useful to pin down the physics of GRBs in the prompt phase.
For example it may reflect the angular structures of the GRB jets. Recently Ghirlanda et al.~(\citeyear{ghir12a})
also connected the bulk Lorentz factors with observable parameters
($E_{\rm peak}, E_{\rm iso}, L_{\rm iso}$) and commendably explained the
$E_{\rm peak}$ -- $E_{\rm iso}$, $E_{\rm peak}$ -- $L_{\rm iso}$ correlations.

The above correlations have prompted GRBs as potential ``standard candles'' in the universe.
Many studies have been devoted to this topic.
Liang et al.~\citeyearpar{liang08} used the SN Ia as a first-order standard candle to calibrate
the GRB correlations, trying to avoid the so called circularity problem (the calibration of GRB
correlations may be cosmology dependent due to the lack of a sufficient low-redshift GRB sample).
Dai et al.~\citeyearpar{dai04} and Amati~\citeyearpar{amati06} have used the Amati relation to
constrain the cosmological models and obtained results consistent with those from the SNe Ia method.
While the reliability of the Amati relation for cosmological use is still under debate (Collazzi et al.~\citeyear{coll12}).
Schaefer~\citeyearpar{schaefer07} argued that the consistency between GRB cosmology
and conventional cosmology is robust, and that further improvements could be expected.
Actually, Wright~\citeyearpar{wright07} have used the GRB data as a supplement to the conventional data of the supernovae,
acoustic oscillations, nucleosynthesis, large-scale structures, and the Hubble constant, to
constrain the dark energy property. He concluded that the GRB Hubble diagram does help to
break the degeneracy between the case of $\omega \neq 1$ and the case of $\Omega_{K} \neq 0$.
However, the large scatter of current GRB relations is still a serious problem
for their applications in cosmology. Tighter or completely new correlations are still the goal
of many researchers (Yonetoku et al.~\citeyear{yone10}; Wang et al.~\citeyear{wang11};
Qi \& Lu~\citeyear{qilu12}; Xu \& Huang~\citeyear{XuHuang12}; Mangano et al.~\citeyear{mang12};
Zhang et al.~\citeyear{zhang12}).

We note from a plot in Amati et al.(2002) that an obvious correlation seems to exist between
$\alpha$ (the low-energy spectra index of the time integrated GRB spectra fitted with the
Band function) and the redshift $z$.  We call this correlation the $\alpha$ -- $z$ relation.
If the $\alpha$ -- $z$ relation really holds, then one can derive the redshifts
of GRBs simply from the observed spectrums. It would be a new and useful method to derive
redshifts, which can be applied in many further studies (Guidorzi~\citeyear{guid05}; Curran et
al.~\citeyear{curran08}). However, the figure in Amati et al. only contains nine GRBs.
No matter whether the $\alpha$ -- $z$ relation is intrinsic or not, it deserves to be tested
by a larger sample. Today, the number of GRBs with known redshifts has been significantly increased
thanks to extensive follow-up observations. In this study, we will
examine the $\alpha$ -- $z$ relation by using a much expanded sample.
We will also explore the correlations between $\alpha$ and other parameters, such as $E_{\rm peak}$,
$E_{\rm iso}$, and $L_{\rm iso}$.

Our paper is organized as follows. In Section 2, we describe the sample used for the study.
In Section 3, the correlations among a variety of parameters are explored. It is found that
there is indeed a correlation between $\alpha$ and $z$, although it is a bit more scattered than the result of
Amati et al. Correlations between $\alpha$ and $E_{\rm peak}$, $E_{\rm iso}$, $L_{\rm iso}$ are also
found. Theoretical implications of these correlations are then explored in Section 4. Finally, Section
5 presents our main conclusions. Possible applications of the $\alpha$ -- $z$ relation are also discussed.

\section{Sample}
\label{sect:sam}

To fit the time-averaged spectra of GRBs, people usually use three kinds of functions: the simple power-law function,
the cutoff power-law function and the Band function (Band et al.~\citeyear{band93}). The simple power-law function is
\begin{equation}
f(E) \propto E^{\rm - \alpha'},
\end{equation}
where $E$ is the photon energy and $\alpha'$ is the power-law index.
The cutoff power-law function is
\begin{equation}
f(E) \propto E^{\rm - \alpha'} \exp(\frac{- E (2 -
\alpha')}{E_{\rm peak}}).
\end{equation}
The Band function is usually expressed as
\begin{equation}
f(E) \propto \left\lbrace \begin{array}{ll} E^{\rm - \alpha}
     \exp(\frac{- E (2 - \alpha)}{E_{\rm peak}}),
     ~~~~~~E < (\frac{- (\alpha - \beta) E_{\rm peak}}{2 - \alpha}), \\
                    \\
    (\frac{- (\alpha - \beta) E_{\rm peak}}{(2 - \alpha)})^{\rm - (\alpha - \beta)} E^{\rm - \beta},
    ~~~~~~E \geq (\frac{- (\alpha - \beta) E_{\rm peak}}{(2 - \alpha)}),
        \end{array} \right.
\end{equation}
where $\alpha$ is the power-law index in the low-energy range, and $\beta$ is the power-law index
in the high-energy range. Note that generally $\alpha$ and $\beta$ are positive in our notation.

Each function has its advantage in specific spectrum fit. The derived spectral parameters
based on different spectral functions are also systematically different. Spectrums of
GRBs in the BATSE catalog are often well fitted by the Band function, while the spectra of GRBs observed by Swift are usually best fitted by
the power-law or the cutoff power-law functions
(Sakamoto et al.~\citeyear{saka11}). The difference may be caused by the relatively
narrow energy response of the Swift/BAT detector (15--150 keV). Band et al. (1993) have
shown that even when the intrinsic spectrum of a burst is a Band function, the observed
spectrum might still be well fitted with a simple power-law function or a cutoff power-law
function if $E_{\rm peak}^{\rm obs}$ is outside the energy range of the detector. This is
intelligible because there is no sufficient data on both sides of $E_{\rm peak}^{\rm obs}$ to credibly constrain
a function with a break (Krimm et al.~\citeyear{krimm09}). So, in order to obtain a good
spectrum of a GRB, the detector should have a wide energy response. A wide energy band also help to get
a good fit result with the Band function. Note that other factors, such as the sensitivity of the
detector (Band~\citeyear{band03}) and the signal-to-noise ratio of the bursts (Krimm et
al.~\citeyear{krimm09}), may also affect the correct derivation
of spectral parameters. A less sensitive instrument or a low signal-to-noise ratio of the burst will
lead to data with bad quality, which could not be used to credibly derive the spectral parameters.
In this study, since we are mainly concentrating on the spectrum features of GRBs, we will only
select those GRBs with high spectrum quality, i.e., they should be observed by a
detector with wide energy response and high sensitivity so that the Band function parameters can
be well constrained from the observations. Additionally, the redshifts of these GRBs also need to be measured.

As discussed above, we need the parameters of the Band function derived from observations for the purpose
of our study. In principle, the sample should be composed of GRBs with known spectroscopic
redshifts and high-quality spectral data so that their spectrums can be well fitted with the
Band function. Many GRBs observed by Swift have redshift data, but unfortunately the narrow energy
band of the Swift/BAT detector (15--150 keV) seriously limits their usage in
our study. As a result, our sample only contains a few GRBs from the Swift catalog. They were
generally observed simultaneously by Swift/BAT and other wide-band detectors such as the
Konus/Wind or Fermi/GBM, so that high-quality spectral data are available.

With the selection criteria mentioned above, we have collected 65 GBRs. The redshifts and spectrum
parameters are listed in Table~1.
For these GRBs, the spectrum are all better fitted with the Band function than with other
functions according to the $\Delta \chi^{\rm 2}$ criterion (an improvement of 6 units
in $\chi^{\rm 2}$ for a change of 1 degree of freedom, Band et al.~\citeyear{band93};
Sakamoto et al.~\citeyear{saka11}). In fact, all the GRBs in our sample have been frequently used
by other authors to test various other correlations
(Ghirlanda et al.~\citeyear{ghir04}; Nava et al.~\citeyear{nava08},~\citeyear{nava11b},~\citeyear{nava12};
Kann et al.~\citeyear{kann10}; Tsutsui et al.~\citeyear{tsut12}). The spectrum parameters of
the same GRB are generally consistent when the event appears in more than one article mentioned
above.

The distributions of $E_{\rm peak}^{\rm obs}$ and $\alpha$ of our sample are shown in Figure~1.
The mean value of $E_{\rm peak}^{\rm obs}$ is 282 keV and the mean value of $\alpha$ is 1.07
(with a standard deviation of $\sigma = 0.324$). These values are similar to those derived from the
BATSE GRBs (Preece et al.~\citeyear{preece00}; Schaefer~\citeyear{schaefer03}; Kaneko et al.~\citeyear{kane06}).
Note that the value of $E_{\rm peak}^{\rm obs}$ is typically well outside the energy band
of the Swift/BAT detector, which clearly demonstrates the inaptness of most Swift GRBs for the
purpose of our study. In Table~1, $E_{\rm iso}$ and $L_{\rm iso}$ are calculated in the energy range
of 1 -- 10$^{\rm 4}$ keV in the GRB rest frame (a scheme proposed by Amati et al. 2002), and k-correction
has been applied in the integration. GRBs are usually classified as long ($T_{\rm90} \geq 2 s$) and short-duration
($T_{\rm90} < 2 s$) categories (Kouveliotou et al.~\citeyear{kouv93}). Under this classification
scheme, all our events are ``long GRBs'' except for GRB 090510.

For our sample, the observational data are from BeppoSAX, Konus/Wind, HETE-2 and Fermi/GBM,
which have broad energy coverage. The sample are heterogeneous in terms of the different
instruments, thus we need to consider the possible effects of the heterogeneity.
Ghirlanda et al.~\citeyearpar{ghir08} analyzed the distributions of GRBs detected by different instruments
(Swift, BATSE, HETE-2, Konus/Wind and BeppoSAX) in the $E_{\rm peak}^{\rm obs}$~--~fluence plane.
They found that the distribution of the heterogeneous sample is not seriously affected by the trigger
sensitivity (the minimum flux to trigger a burst). Except for the Swift sample (luckily not included
in our sample), the distribution is also not affected by different spectral threshold (the minimum fluence
required to get the spectrum and constrain the peak energy). For Fermi/GBM and BATSE long GRBs, they have
similar distributions of fluence, $E_{\rm peak}^{\rm obs}$, and peak flux. It is also found that the
Fermi/GBM bursts generally have harder low-energy spectral indices ($\alpha$) with respect to the BATSE GRBs,
but the difference is very slight (Nava et al.~\citeyear{nava11a}). So, although the heterogeneity of
the bursts in our sample might do lead to some complex bias, it should not be too serious in general.

\section{Correlations}
\label{sect:corre}

We first investigate the $\alpha$ -- $z$ relation mentioned by Amati et al.(2002) with our
larger sample. The original $\alpha$ -- $z$ relation in previous article is
$ \log \alpha = (-0.78 \pm 0.13) \log (1 + z) + (0.39 \pm 0.04)$.
In Figure~2, we plot our sample in the $\log \alpha - \log (1 + z)$ plane. It
confirms the decreasing trend of $\alpha$ with $z$, though the data points seem dispersive.
Also, the slope of this correlation becomes flatter when the sample is expanded.
The best fit result of the correlation is now
\begin{equation}
\log \alpha = (-0.42 \pm 0.07) \log (1 + z) + (0.11 \pm 0.02).
\end{equation}
The correlation coefficient is $r^{\rm 2} = 0.35$.
The associated p-value (the probability that such a correlation is simply formed by chance) is $2.3 \times 10^{-7}$.
In Figure~3, we plot our sample in the
$\log \beta - \log (1 + z)$ and $\alpha - \beta$ planes. Massaro et al. once suggested that
the $\alpha$ -- $z$ relation could be a manifestation of the selection effect:
brighter GRBs, which can potentially be detected at higher redshifts, should be softened
due to the expansion of the Universe, thus they might have a flatter photon index
(Massaro et al.~\citeyear{massa02}). If this explanation
is correct, then both $\alpha$ and $\beta$ should have similar correlation features with
$z$, and additionally, $\alpha$ and $\beta$ themselves should be correlated.
However, Figure~3 shows clearly that no correlation exists between $\beta$ and $z$.
Further more, the right panel of Figure~3 also shows that there is no positive correlation between
$\alpha$ and $\beta$.

In fact, from theoretical aspect, the power-law spectral indices of the Band function should not
be affected by cosmological redshifting. Assuming that the spectrums at rest frame and observer frame
have similar function form as in Eq. (3), and denoting the spectral parameters in the rest frame and
observe frame with a subscript ``rest'' and ``obs'', then according to the conservation of photons
in a given unit energy interval, we have
\begin{equation}
\frac{{\rm d}N_{\rm rest}(E)}{{\rm d} E}{{\rm d} E} =
    \frac{{\rm d}N_{\rm obs}(\frac{E}{1 + z})}{{\rm d} E}{{\rm d} (\frac{E}{1 + z})}.
\end{equation}
This equation can be re-written as
\begin{equation}
E^{\rm - \alpha_{\rm rest}} \exp(\frac{- E (2 - \alpha_{\rm rest})}{E_{\rm peak}^{\rm rest}})
   \propto (\frac{1}{1 + z})^{\rm 1 - \alpha_{\rm obs}} E^{\rm - \alpha_{\rm obs}}
   \exp(\frac{- E (2 - \alpha_{\rm obs})}{E_{\rm peak}^{\rm rest}}),
\end{equation}
where $E_{\rm peak}^{\rm rest} = (1 + z) E_{\rm peak}^{\rm obs}$. Eq. (6) leads to
$\alpha_{\rm rest} = \alpha_{\rm obs}$. It indicates that the cosmological redshifting does not modify
the value of the power-law index. So the $\alpha$ -- $z$ correlation should not be due to the
reddening of photons. Actually, on the contrary, the $\alpha$ -- $z$ relation revealed in
Figure~2 indicates that the low-energy spectrum gets harder with the increasing $z$.
Additionally, the facts that $\beta$ does not correlate similarly with $z$, and that
$\beta$ does not positively correlate with $\alpha$ (Figure~3) also strongly
indicate that the $\alpha$ -- $z$ correlation should be due to some intrinsic
mechanism.

However, it should be noted that if $E_{\rm peak}^{\rm obs}$ is shifted to be near
the low-energy limit of the passband, then the low-energy data may not be sufficient enough
to give the correct spectrum index. This kind of bias could be presented in the value of $\alpha$.
For example, a simple extrapolation of the Band spectra from the Fermi/GBM band to the Fermi/LAT band would
systematically over-predict the observed flux (Ackermann et al.~\citeyear{acker12}). This can
be explained by the softer $\beta$ values or intrinsic spectral breaks at energies
$\geq 40 {\rm MeV}$, a similar effect as discussed here.
In Figure~4, we show our sample on the $E_{\rm peak}^{\rm obs}$ -- $(1 + z)$
plane to see to what extent this kind of bias might affect our results. Luckily, for most detectors,
the lower limits are below 10 keV. As a result, actually all the GRBs lie significantly above the
corresponding detector limit. So our results would not be seriously affected by this bias.

Sakamoto et al.~\citeyearpar{saka09} have found an
empirical relation between the power-law spectum indices of the Swift/BAT GRBs ($\Gamma^{\rm BAT}$)
and $E_{\rm peak}$: $\log E_{\rm peak}^{\rm obs} \propto \Gamma^{\rm BAT}$. In Figure~5,
$\alpha$ vs $E_{\rm peak}$ and $\beta$ vs $E_{\rm peak}$ are plotted for the GRBs in our
sample. We find that an obvious correlation exists between $\alpha$ and $E_{\rm peak}$,
which can be best fitted as,
\begin{equation}
\log \alpha = (-0.14 \pm 0.03) \log E_{\rm peak} {\rm (keV)} + (0.33 \pm 0.07).
\end{equation}
The corresponding correlation coefficient is $r^{\rm 2} = 0.30$ and the p-value is
$4.3 \times 10^{-6}$. We call this relation the $\alpha$ -- $E_{\rm peak}$ relation. It
implies that a GRB with a larger $E_{\rm peak}$ tends to have a harder low-energy spectrum.
However, we should notice that $E_{\rm peak}$ itself is mathematically related to alpha
through the definition of the e-folding parameter, $E_{0}$, by $E_{\rm peak} = E_{0} \ast (2 - \alpha)$.
Therefore, the $\alpha$ -- $E_{\rm peak}$ relation may be partially due to the definition.
No correlation exists between $\beta$ and $E_{\rm peak}$,
as shown in the right panel of Figure~5.

We have studied the correlation between the spectral parameters and the total radiated energy.
In Figure~6, we plot the sample in the $\log \alpha$ -- $\log E_{\rm iso}$ plane. As clearly shown in
the left panel of Figure~6, we see evidence of a correlation between $E_{\rm iso}$ and
$\alpha$ (called the $\alpha$ -- $E_{\rm iso}$ relation hereafter). The best fit result is
\begin{equation}
\log \alpha = (-0.08 \pm 0.01) \log E_{\rm iso} {\rm (erg)} + (4.3 \pm 0.55).
\end{equation}
The corresponding correlation coefficient is $r^{\rm 2} = 0.53$ and the p-value is $3.8 \times 10^{-10}$.
It implies that the more energetic
one GRB be, the harder its low-energy spectrum is. But interestingly, for the high-energy
index ($\beta$), no trend of correlating with $E_{\rm iso}$ can be found (see the right panel
of Figure~6).

We have also investigated the correlation between the spectral parameters and the isotropic
luminosity. The results are plotted in Figure~7. The left panel of Figure~7 shows that
$\alpha$ correlates with $L_{\rm iso}$ tightly. The best fitted function is
\begin{equation}
\log \alpha = (-0.1 \pm 0.01) \log L_{\rm iso} {\rm (erg/s)} + (5.2 \pm 0.6).
\end{equation}
The corresponding correlation coefficient is $r^{\rm 2} = 0.63$ and the p-value is $3.8 \times 10^{-11}$.
Such an $\alpha$ -- $L_{\rm iso}$ relation
is even tighter than the $\alpha$ -- $E_{\rm iso}$ relation displayed in Figure~6.
On the other hand, no correlation can be found between $\beta$ and $L_{\rm iso}$, as shown in the
right panel of Figure~7.

Using our expanded sample, we can also examine the Amati relation. The results are plotted in Figure~8.
From the left panel, we see that $E_{\rm peak}$ and $E_{\rm iso}$ are highly correlated. Our best fit
result is:
\begin{equation}
\log E_{\rm peak} {\rm (keV)} = (0.46 \pm 0.04) \log E_{\rm iso} {\rm (erg)} - (22 \pm 2.2).
\end{equation}
The corresponding correlation coefficient is $r^{\rm 2} = 0.71$ and the p-value is $2.6 \times 10^{-15}$.
The slope is $\sim 0.5$, consistent with
previous studies. In the right panel of Figure~8, we plot the diagram of $L_{\rm iso}$ vs
$z$. It is shown that there is a tendency that the luminosity increases with increasing $z$.
This could be due to the selection effect that faint GRBs are difficult to be detected at higher
redshifts. The evolution of $L_{\rm iso}$ with respect to $z$ can be fitted with a
power-law function of
\begin{equation}
\log L_{\rm iso} {\rm (erg/s)} = (2.56 \pm 0.47) \log (1 + z) + (51.26 \pm 0.14),
\end{equation}
for which the correlation coefficient is $r^{\rm 2} = 0.39$ and the p-value is $2.3 \times 10^{-6}$.

For our sample, there is also a clear correlation between $E_{\rm peak}$ and $z$.
Figure~9 shows that for more distant GRBs, the $E_{\rm peak}$ value is generally higher.
But generally speaking, the correlation is not so tight, roughly consistent with previous studies
(Mallozzi et al.~\citeyear{mallo95}; Wei \& Gao~\citeyear{weigao03}).

As a comparison, we have also selected all the
Swift GRBs\footnote{http://swift.gsfc.nasa.gov/docs/swift/archive/grb\_table/} with known
redshifts, trying to examine any possible correlations between the spectral parameters and the
redshifts. For the majority of these Swift GRBs, the spectral parameters of the Band function are
not available due to the narrow energy range of the Swift/BAT detector. Their spectrums can
only be fitted with a power-law or a cutoff power-law function. For these events, the derived
power-law spectral indices may somewhat be similar to the lower-energy spectral indices
of the Band function. However, somewhat unexpectedly, we find that there is no correlation
between the power-law index and the redshift. It may indicate that the narrow energy
range of the Swift/BAT detector has seriously limited our determination of the spectrum,
even at the low-energy segment.

\section{Physical Implications}
\label{sect:physi}

For the synchrotron emission from shocked electrons in magnetic fields (Rybicki
\& Lightman~\citeyear{rybi79}), the low-energy spectrum index $\alpha$ should range from $2/3$,
in the case of optically thin synchrotron emission (Katz~\citeyear{katz94}), to $3/2$, when
the distribution of electrons is characterized by a power-law index of $-2$ (Preece et
al.~\citeyear{pree98}). However, a notable fraction of the observed GRB spectra
are harder than the optically thin synchrotron spectra (i.e. with $\alpha < 2/3$). This is
the so-called ``death line'' problem (Preece~\citeyear{pree98}).

Thermal emission may provide a solution to the ``death line'' problem.
Ryde~\citeyearpar{ryde05} analyzed a sample of 25 pulses in the catalog of
Kocevski et al.~\citeyearpar{koce03}. He found that thermal emission, combined with a
nonthermal component, is ubiquitous for GRBs of various spectral shapes and timing
characters. So, the GRB radiation may consist of both blackbody emission from
photosphere and nonthermal component from relativistic outflows (Ryde 2004; Fan et al.~\citeyear{fan12}).
The measured value of $\alpha$ then should depend on the relative strength of the thermal
component and the nonthermal component. Since the Rayleigh-Jeans portion of a blackbody
spectrum is a power law function with $\alpha = - 1$, it may lead $\alpha$ to exceed the synchrotron
range. For example, if the thermal component is strong enough to become the dominant component,
then the resulted spectrum will have a small $\alpha$.

Strong photospheric emission at gamma-ray wavelengths is predicted in most GRB scenarios.
In usual outflow models (Piran~\citeyear{piran99}; M{\'e}sz{\'a}ros~\citeyear{mesz02};
Piran~\citeyear{piran04}; Rees \& M{\'e}sz{\'a}ros~\citeyear{rees05}),
the typical temperature of the photosphere is $T \propto L^{\rm 1/4}$.
It indicates that a higher luminosity usually corresponds to a stronger thermal component,
which then leads to a smaller $\alpha$. This may naturally give an
explanation to the $\alpha$ -- $L_{\rm iso}$ correlation, as displayed in Figure~7.
At the same time, since both $L_{\rm iso}$ and $E_{\rm iso}$ are indicative of the strength of
GRBs, it is easy to understand that these two parameters should be positively correlated.
Combining the $\alpha$ -- $L_{\rm iso}$ correlation and the $L_{\rm iso}$ -- $E_{\rm iso}$
correlation, we can then easily explain the $\alpha$ -- $E_{\rm iso}$ correlation.

In Figure~10, we illustrate the relationship of all the correlations studied in our article.
On the left side, from the $\alpha$ -- $E_{\rm iso}$ correlation (Figure~6) and the Amati relation
(i.e. the $E_{\rm peak}$ -- $E_{\rm iso}$ correlation, Figure~8), we can get the $\alpha$ --
$E_{\rm peak}$ correlation
(see Figure~5, of course, the effect of definition as mentioned before
cannot be ignored either). On the right side, from the $\alpha$ -- $L_{\rm iso}$ correlation
(Figure~7) and the $L_{\rm iso}$ -- $z$ correlation (a selection effect, as discussed in the previous
section, also see Figure~8), we can get the $\alpha$ -- $z$ correlation (Figure~2). Finally,
from the $\alpha$ -- $E_{\rm peak}$ and $\alpha$ -- $z$ correlations, we then have the
$E_{\rm peak}$ -- $z$ correlation (Figure~9).

However, we must bear in mind that the above analysis is only qualitative. When we check the
correlations in detail, we get some difference. For example, Eq.~(9) (i.e. the fit result
of Figure~7) indicates that $ \log \alpha \propto -0.1 \log L_{\rm iso}$, and Eq.~(11) (i.e. the
fit result of the left panel of Figure~8) indicates $\log L_{\rm iso} \propto 2.56 \log (1 + z)$.
Combining these two equations, we then get $\log \alpha \propto -0.26 \log (1 + z)$.
However, such a result is slightly different from the best fit result of Figure~2,
which actually gives $\log \alpha \propto -0.42 \log (1 + z)$. The difference indicates
that there might be other factors that are playing their roles. They need to be studied
further in the future.

The largest uniform sample of GRBs with measured low-energy spectral indices (i.e. the
$\alpha$ parameter of the Band function) should be the BATSE GRBs. However, the redshifts
of most of the BATSE GRBs are unknown. Fortunately, we could estimate the redshifts from
some statistics relations. We have calculated the pseudo redshifts of the
BATSE GRBs by using the so called Yonetoku relation (the $E_{\rm peak}$ -- $L_{\rm iso}$ correlation,
Yonetoku et al. 2004). We then can plot the BATSE GRBs on the $\log \alpha - \log (1 + z)$ plane.
The results are shown in Figure~11. In this figure, we have also plotted our sample as the
red dots.
It is interesting to see that the BATSE GRBs also show a correlation between
$\alpha$ and $z$, i.e. $\log \alpha \propto -0.23 \log (1 + z)$. The behavior of BATSE GRBs and
our sample are generally similar, though the slope is slightly different.
It indicates that the $\alpha$ -- $z$ correlation studied here may also hopefully act as a new
redshift indicator.

\section{Discussion and Conclusions}
\label{sect:disc}

We have collected a large sample of 65 GRBs from the literature, of which both the redshifts and
the Band function spectral parameters are available. Using this greatly expanded sample,
we investigated the correlation between $\alpha$ and $z$, as first proposed by Amati in 2002.
It is confirmed that the $\alpha$ -- $z$ correlation does exist, although it seems more
scattered in our case. At the same time, it is interesting to note that the high-energy spectral
index, $\beta$, does not correlate with $z$. A few other correlations are also found
to exist for our sample, including the $E_{\rm peak}$ -- $E_{\rm iso}$, $\alpha$ -- $E_{\rm iso}$,
$\alpha$ -- $L_{\rm iso}$, $L_{\rm iso}$ -- $z$, $\alpha$ -- $E_{\rm peak}$, $E_{\rm peak}$ -- $z$ correlations,
etc. Among all these correlations, the $\alpha$ -- $L_{\rm iso}$ relation is the tightest.
Again, although the low-energy spectral index ($\alpha$) shows an obvious correlation with
many parameters such as $E_{\rm iso}$, $L_{\rm iso}$, $z$, $E_{\rm peak}$, the high-energy spectral
index ($\beta$) does not correlate with any of these parameters.

The $\alpha$ -- $z$ correlation indicates that the ``death line'' problem is more serious
at high redshifts. We have shown that this correlation can not be simply attributed to the
photon reddening induced by cosmological expansion. Amati (2002) suggested that the
correlation may hint that the radiative cooling occurs more actively in GRBs at smaller
redshift. However, the intrinsic factor that leads to such a difference of cooling is
completely uncertain. According to the statistics of Ryde, thermal emission may be a common
component in GRBs. Here we prefer to use the thermal component assumption to explain the
correlation. For GRBs with a higher luminosity (i.e., a larger $L_{\rm iso}$), the relative
strength of the thermal component with respect to the non-thermal component is also
higher, leading to a smaller $\alpha$. At the same time, at higher redshifts, only those
GRBs with a relatively higher luminosity are likely to be detected (i.e., the $L_{\rm iso}$ -- $z$
correlation, as displayed in Figure~8, which is actually a reasonable selection effect).
Considering these two factors, the $\alpha$ -- $z$ correlation can then be naturally explained.
Based on this idea, the other correlations mentioned above can all be easily understood (see
the schematic illustration in Figure~10).

For the BATSE GRBs, we have calculated their pseudo redshifts by using the Yonetoku relation
(Yonetoku et al. 2004). When plotted on the $\log \alpha$ -- $\log (1+z)$ diagram, it is
found that the BATSE GRBs show similar tendency of correlation as our sample. This indicates
that the $\alpha$ -- $z$ correlation might serve as a new redshift indicator in the future. When
the Band function of the spectrum of a GRB is available, the low-energy spectral index might
be used to give useful information on the redshift.

In many studies, when the spectrums of GRBs are involved, it is usually assumed that the
intrinsic spectrum shape is a standard one, or at least it does not evolve with redshift.
For example, Ghirlanda et al.~\citeyearpar{ghir12b} simulated a population of bursts to
inspect the impact of selection biases on the $E_{\rm peak}$ -- $L_{\rm iso}$ correlation of GRBs.
They assumed that the GRB spectra has fixed low and high energy spectral indices over
the range of redshift investigated. But according to our current study, the observed
spectral parameter $\alpha$ actually evolves with redshift. This may either be due to
the fact that the intrinsic $\alpha$ correlates with other parameters such as $L_{\rm iso}$
or $E_{\rm iso}$, or that the intrinsic $\alpha$ itself evolves with redshift. In any case,
there does not exist a standard shape for the intrinsic spectrum.

GRBs can be approximately divided into two categories, short GRBs with the duration less than
$\sim$~2 seconds and long GRBs with the duration larger than $\sim$~2 seconds.
It is generally
believed that short GRBs may be produced by the merging of compact binaries, while long GRBs
come from the collapse of massive stars (Woosley~\citeyear{woos93}; Paczynski~\citeyear{paczy98};
MacFadyen \& Woosley~\citeyear{macwoos99}).
It should be noted that most of the GRBs in our sample
are long GRBs. The correlations investigated here thus are only applicable to long GRBs. It is
an interesting problem that whether these correlations also exist for short GRBs. The problem
should be solved when more and more short GRBs with measured redshifts and well determined
spectrums are available.

\acknowledgments
We thank the anonymous referee for valuable comments and suggestions.
This work was supported by the National Basic Research Program of China (973 Program, Grant
No. 2009CB824800) and the National Natural Science Foundation of China (Grant Nos. 11033002 and J1210039).

\appendix

\clearpage

\begin{figure}
   \plottwo{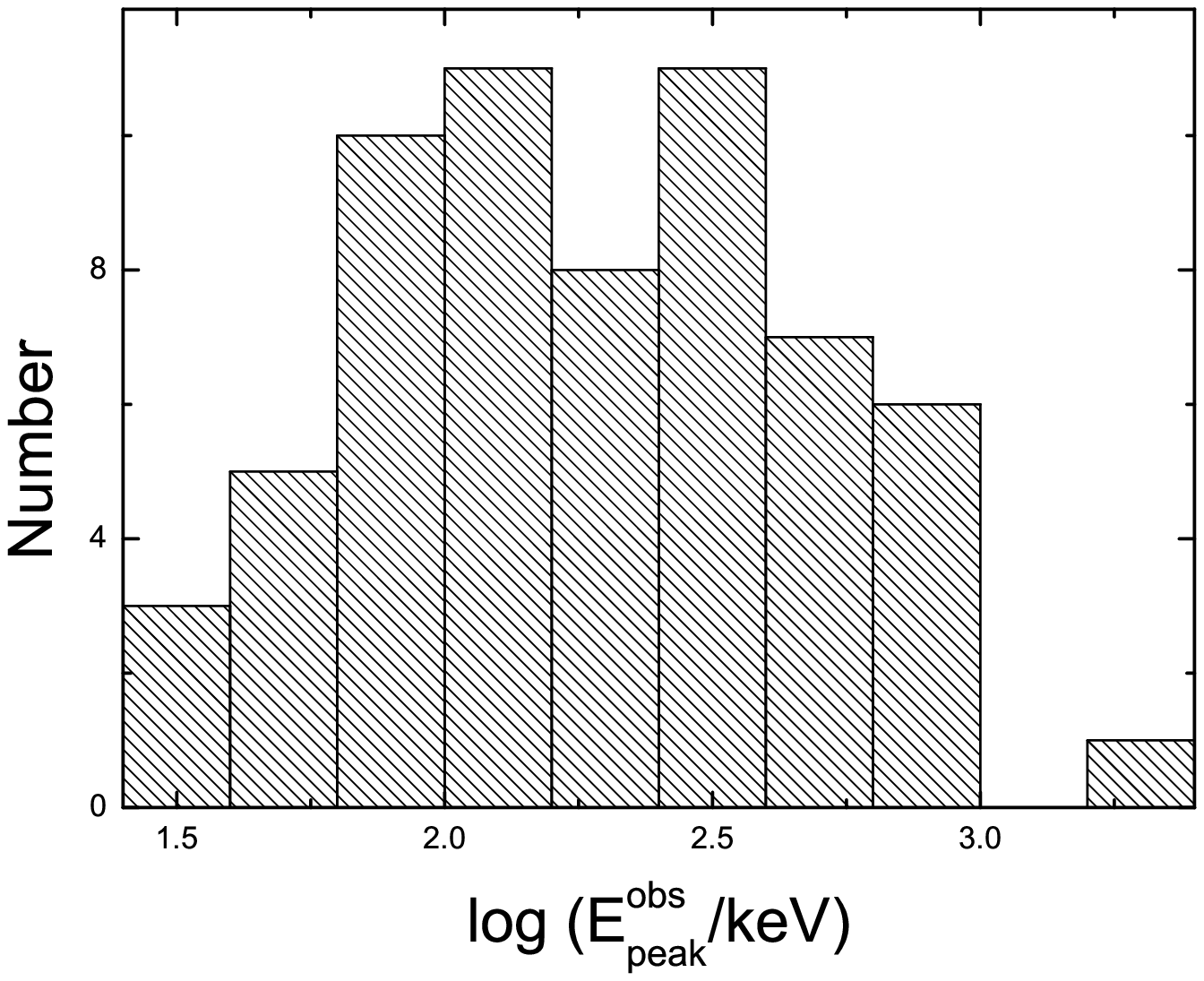}{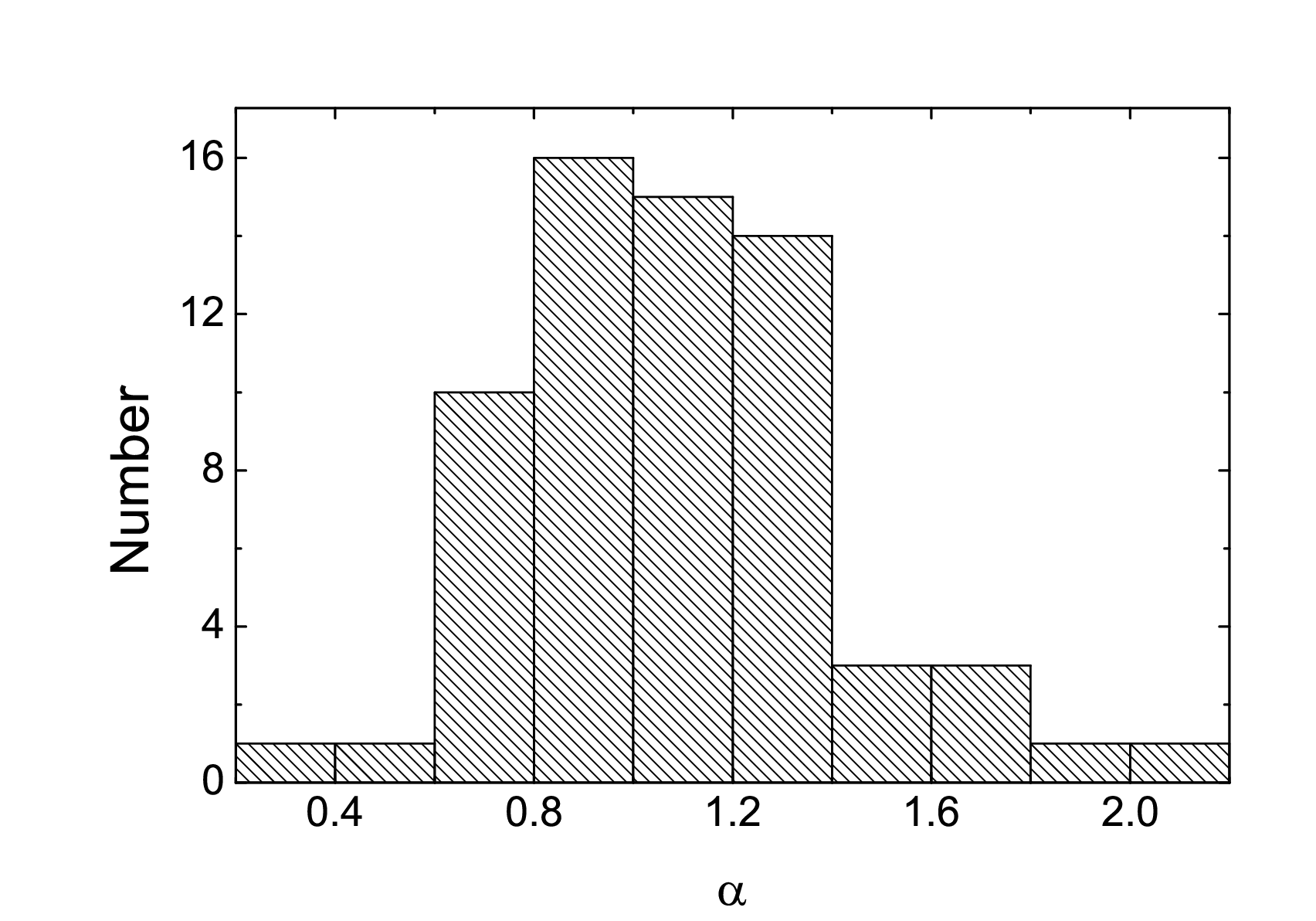}
   \caption{Left panel: distribution of $E_{\rm peak}^{\rm obs}$ of our sample.
Right panel: distribution of the low-energy spectral index ($\alpha$) of our sample.}
   \label{Fig:plot1}
\end{figure}

\begin{figure}
   \vspace{0.5cm}
   \begin{center}
   \plotone{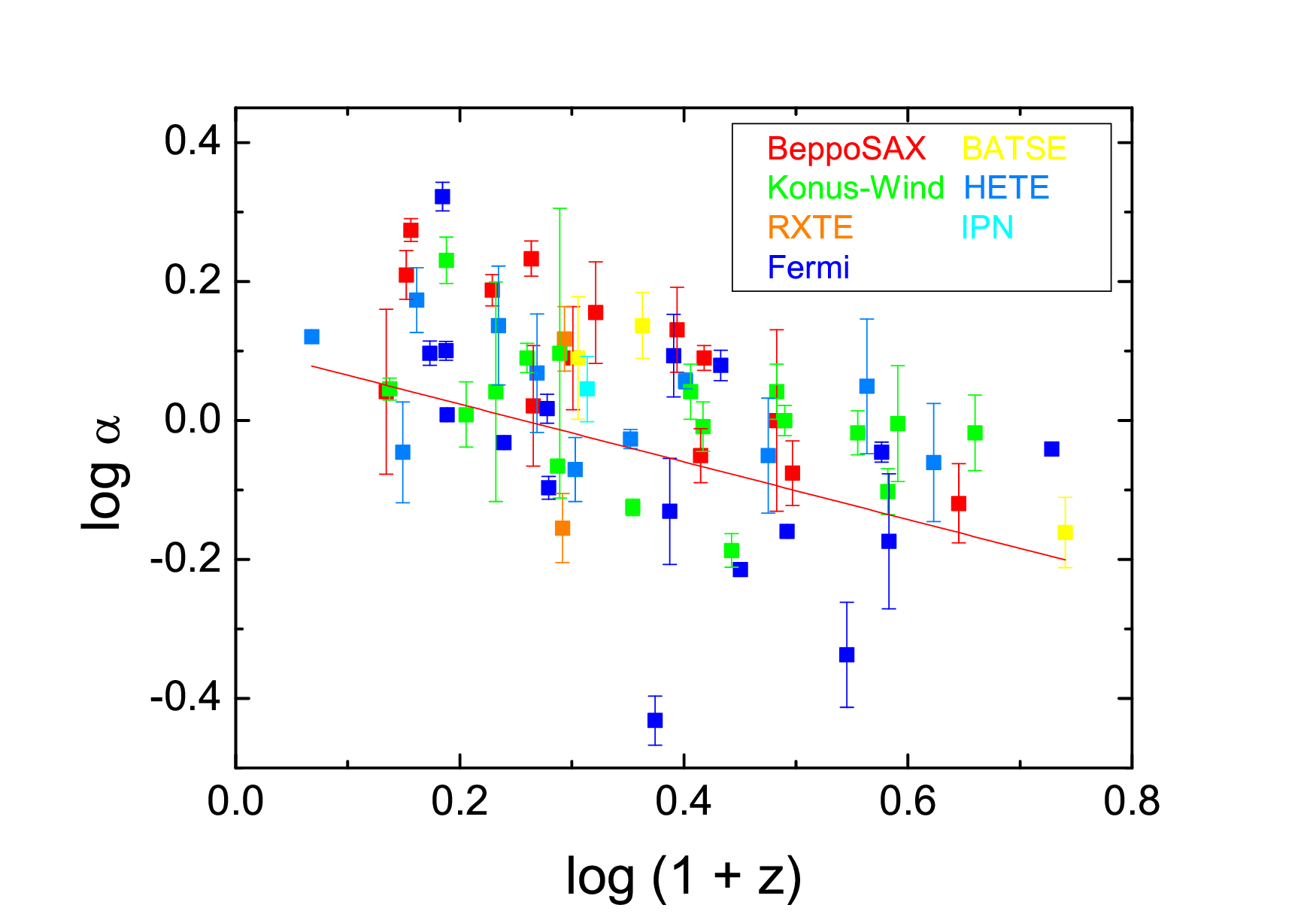}
   \caption{Correlation between the low-energy spectral index and the redshift for our
GRB sample. The solid line is our best fit.}
   \label{Fig:plot2}
   \end{center}
\end{figure}

\begin{figure}
   \plottwo{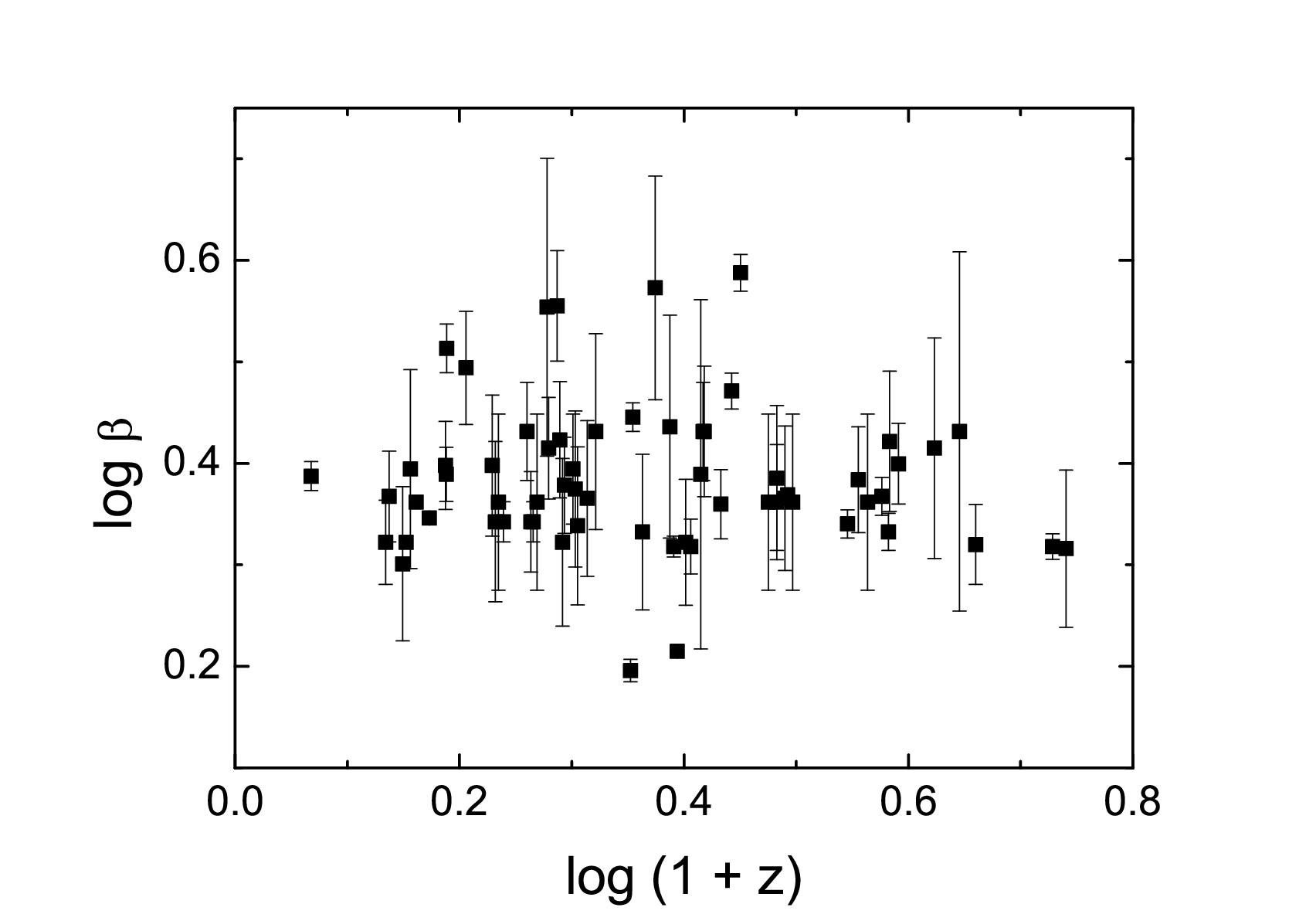}{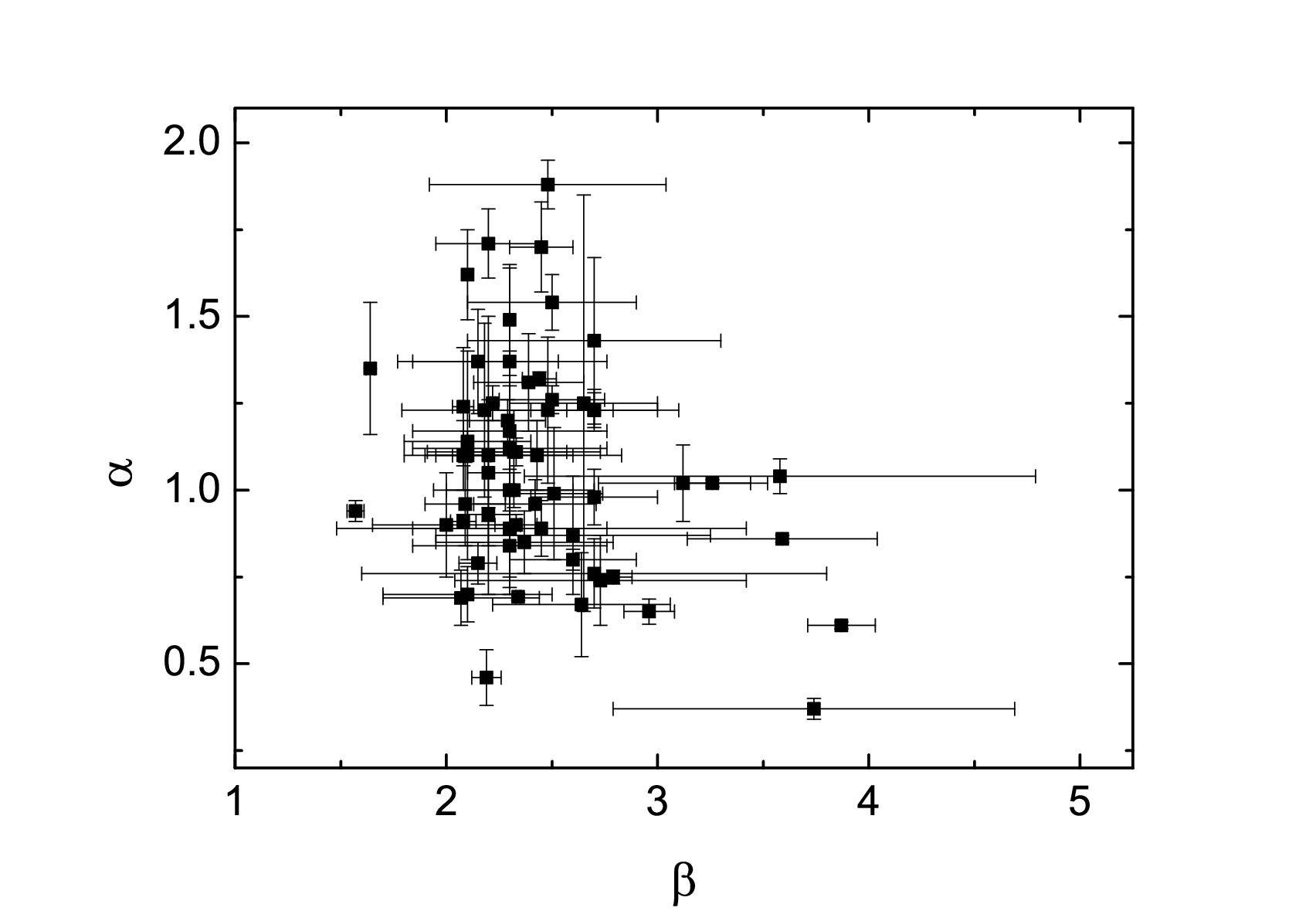}
   \caption{Left panel: dependence of the high-energy spectral indice ($\beta$) on redshift
            for the GRBs in our sample (GRB 081007 is not plotted since its
            error bar is too large). Right panel: $\alpha$ vs $\beta$ for the same GRBs. }
   \label{Fig:plot3}
\end{figure}

\begin{figure}
   \vspace{0.5cm}
   \begin{center}
   \plotone{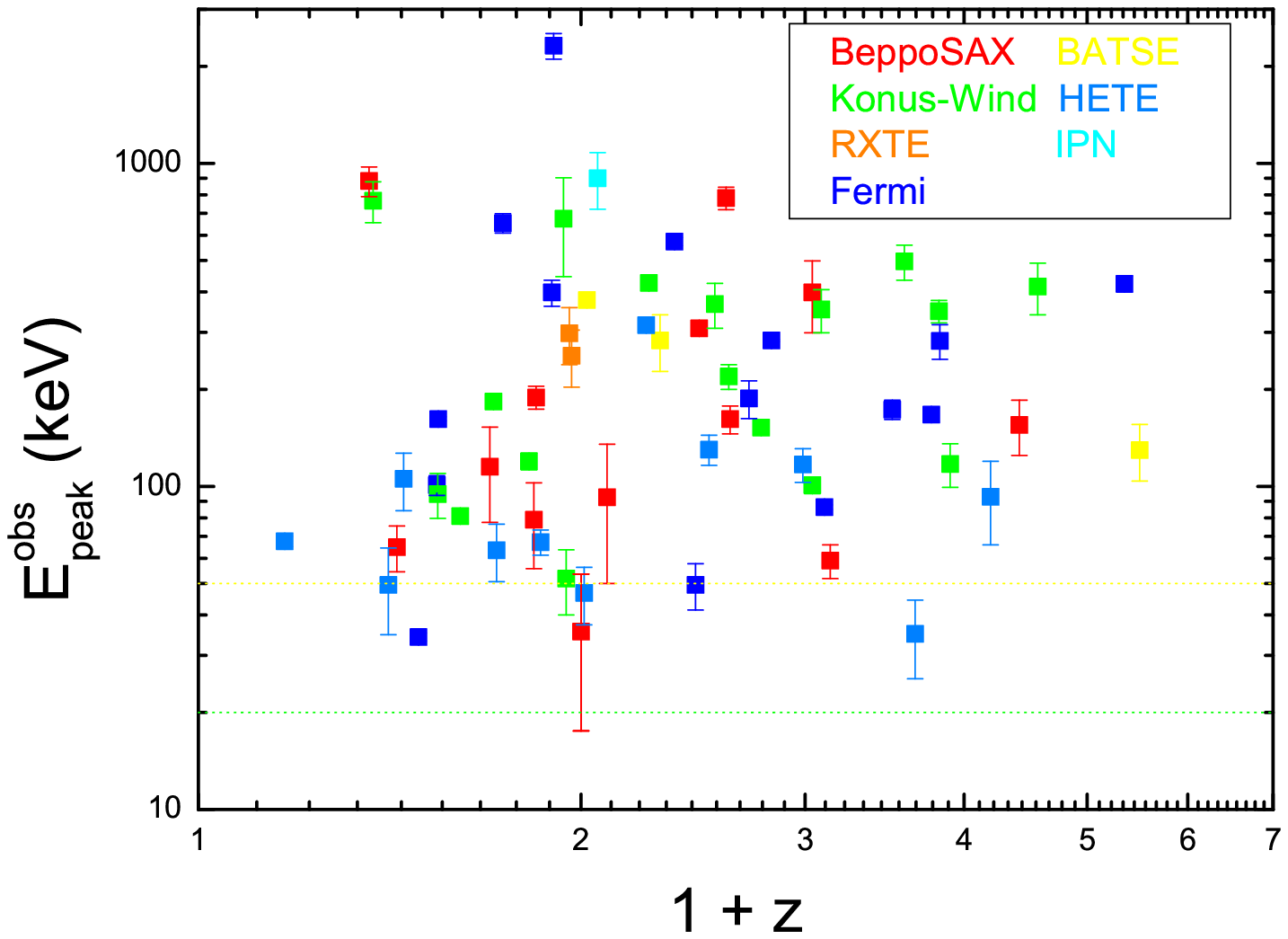}
   \caption{Distribution of the GRBs of our sample on the $E_{\rm peak}^{\rm obs}$ --
$(1 + z)$ plane. The dotted lines correspond to the lower enegy limits of BATSE (yellow) and
Konus/Wind (green) detectors. The limits of other detectors are not shown here because
they are below 10 ${\rm keV}$. }
   \label{Fig:plot4}
   \end{center}
\end{figure}

\begin{figure}
   \plottwo{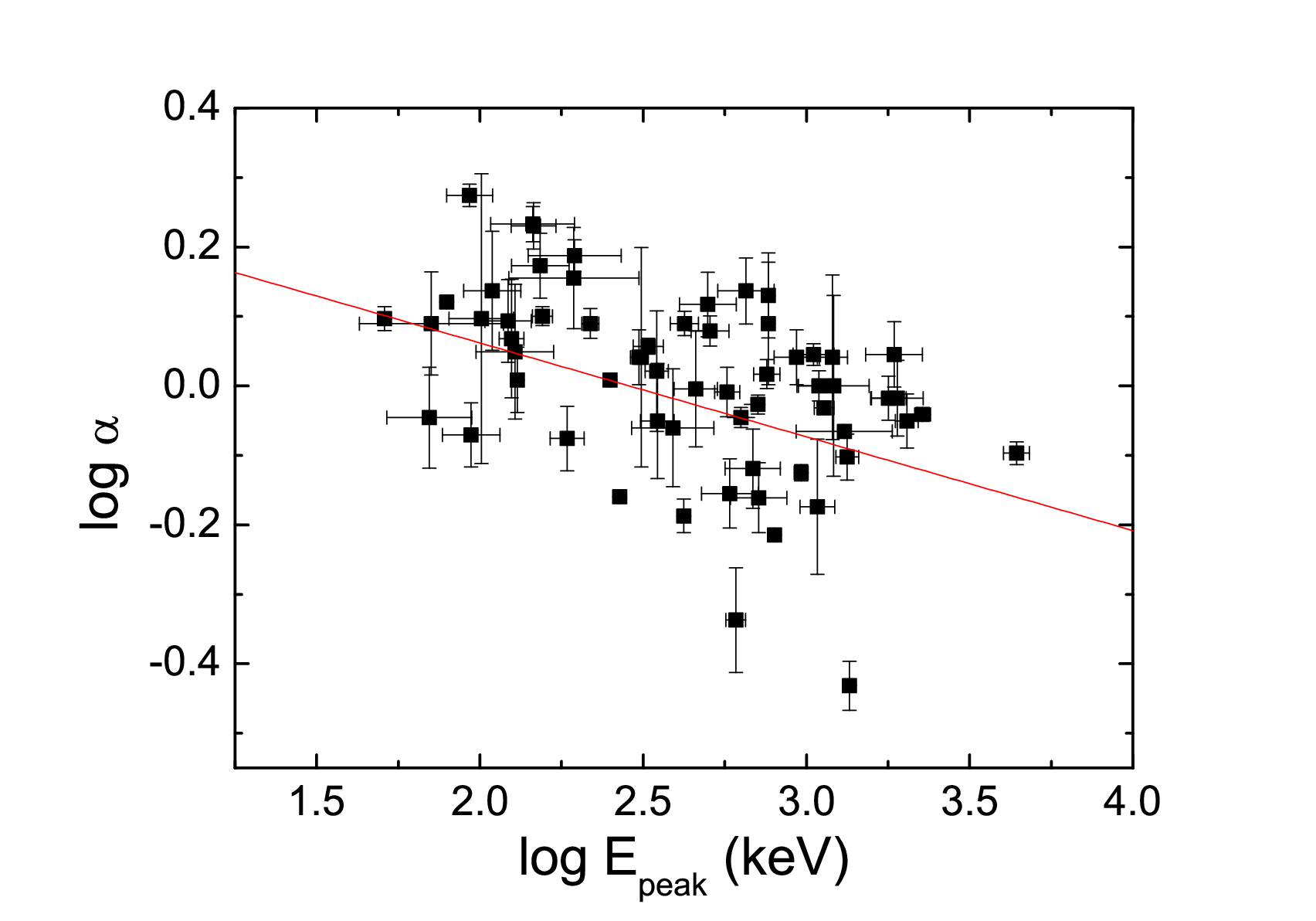}{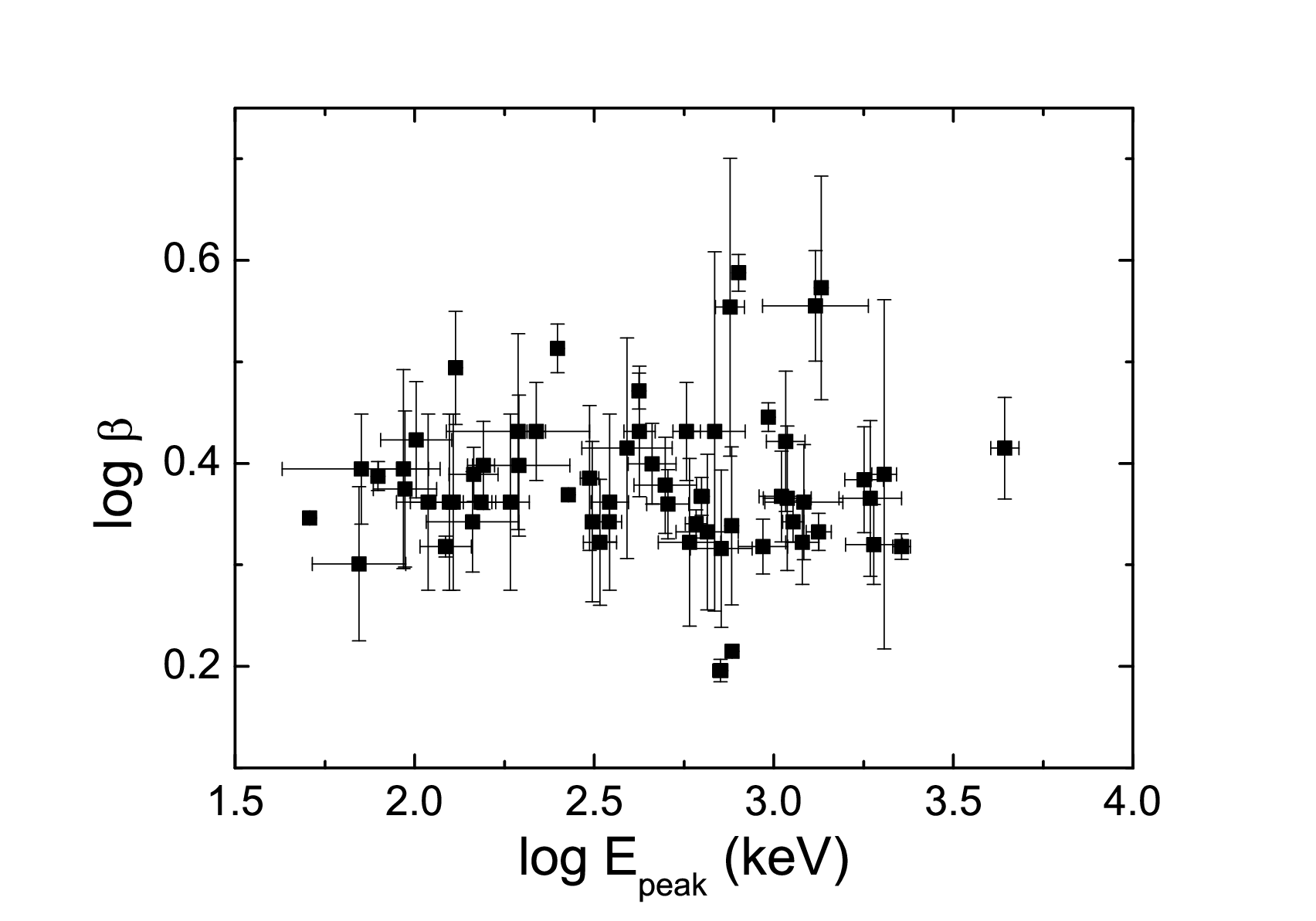}
   \caption{Left panel: $\alpha$ vs $E_{\rm peak}$ for the GRBs in our sample. The solid line
            corresponds to the best fit. Right panel: the corresponding $\beta$ vs
            $E_{\rm peak}$ diagram. }
   \label{Fig:plot5}
\end{figure}

\begin{figure}
   \plottwo{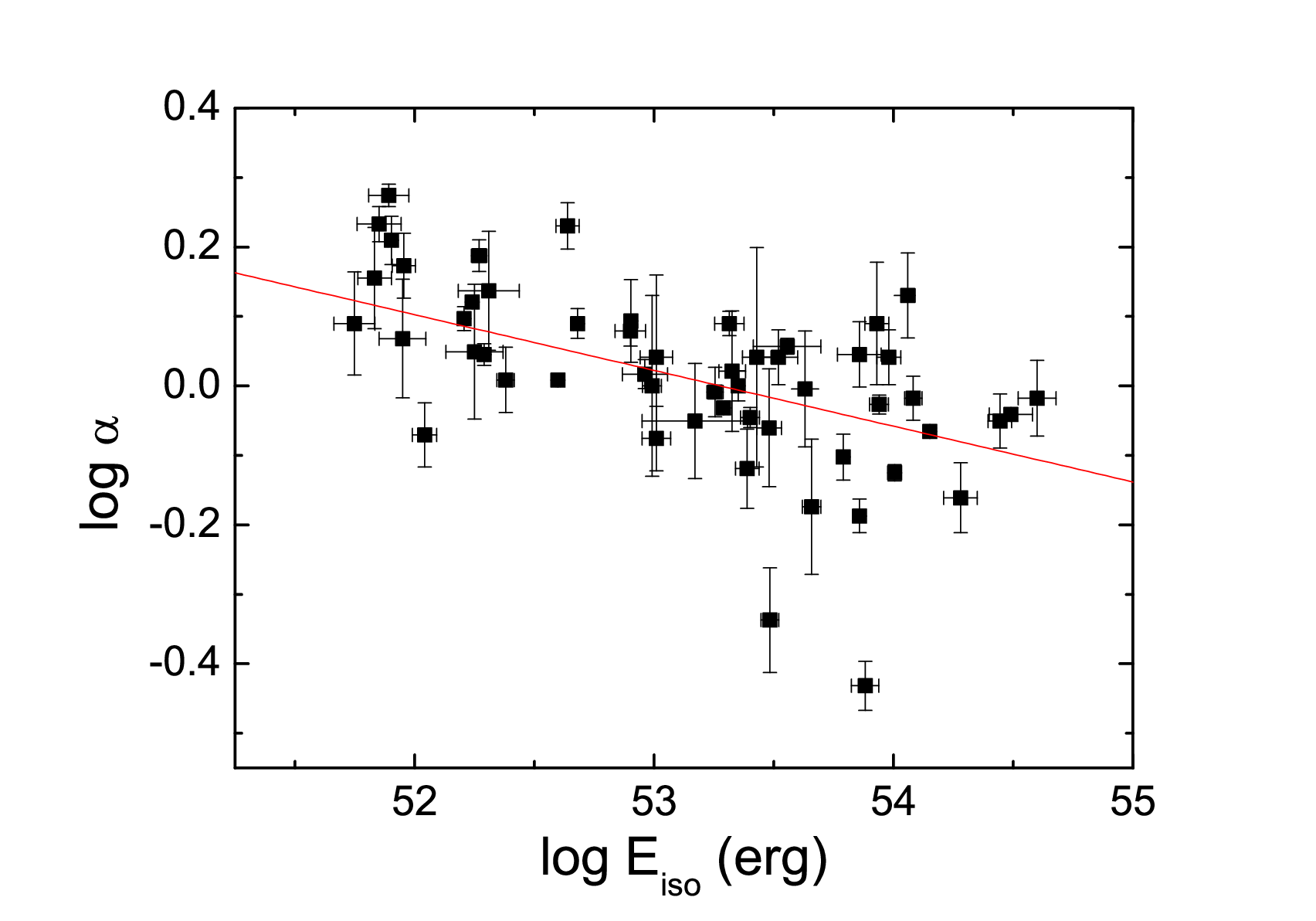}{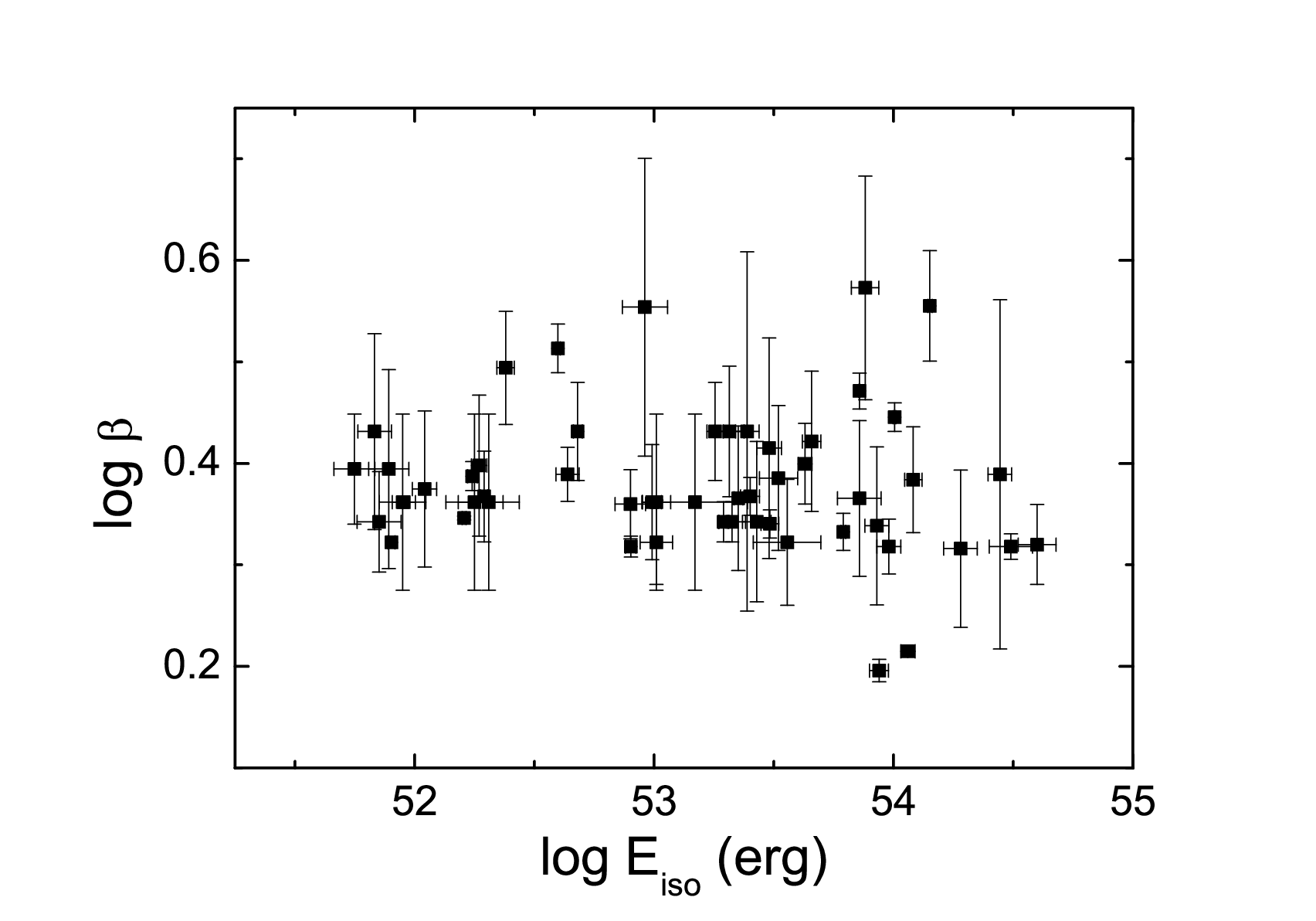}
   \caption{Left panel: $\alpha$ vs $E_{\rm iso}$ for the GRBs in our sample. The solid line
            corresponds to the best fit. Right panel: the corresponding $\beta$ vs
            $E_{\rm iso}$ diagram. }
   \label{Fig:plot6}
\end{figure}

\begin{figure}
   \plottwo{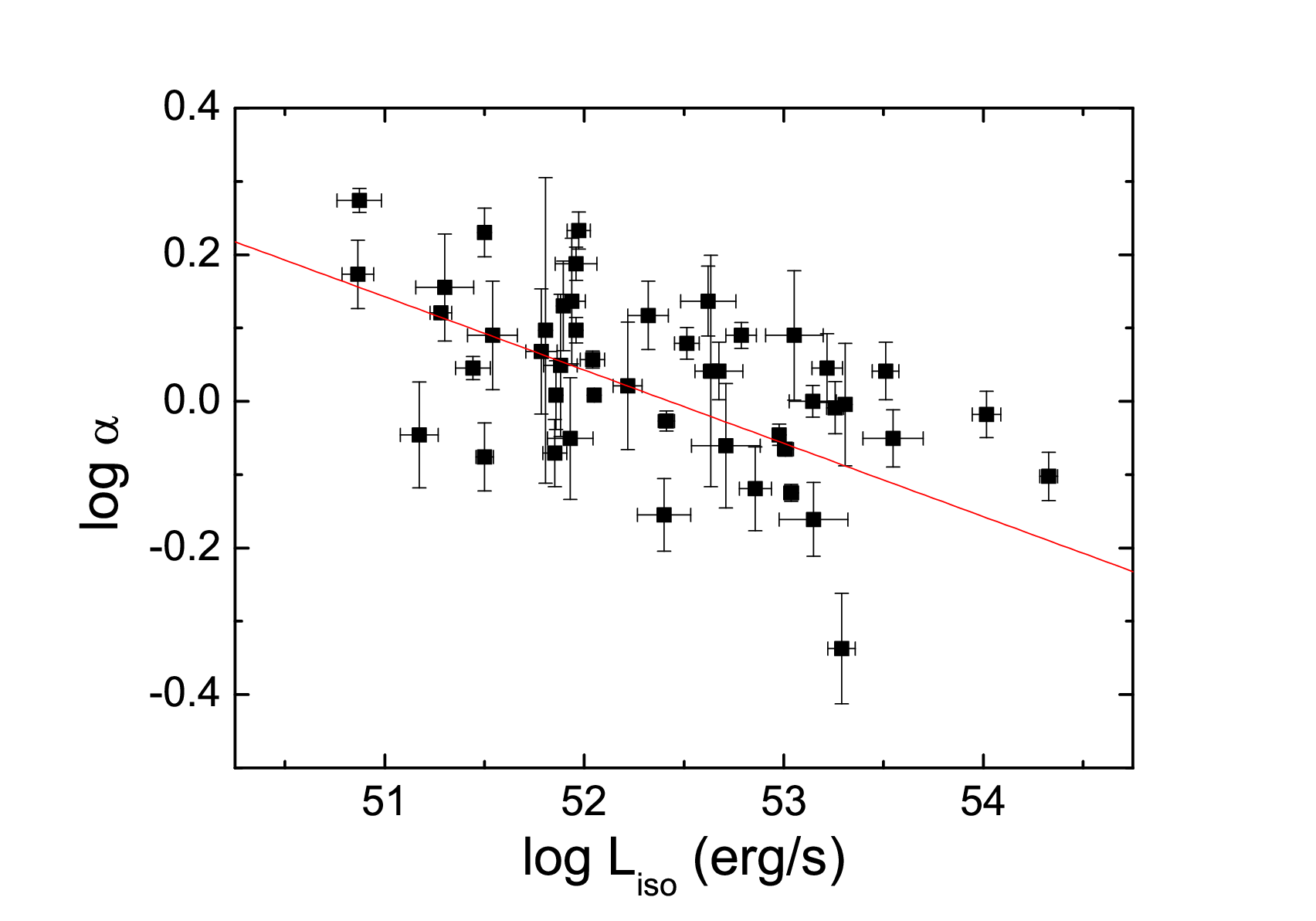}{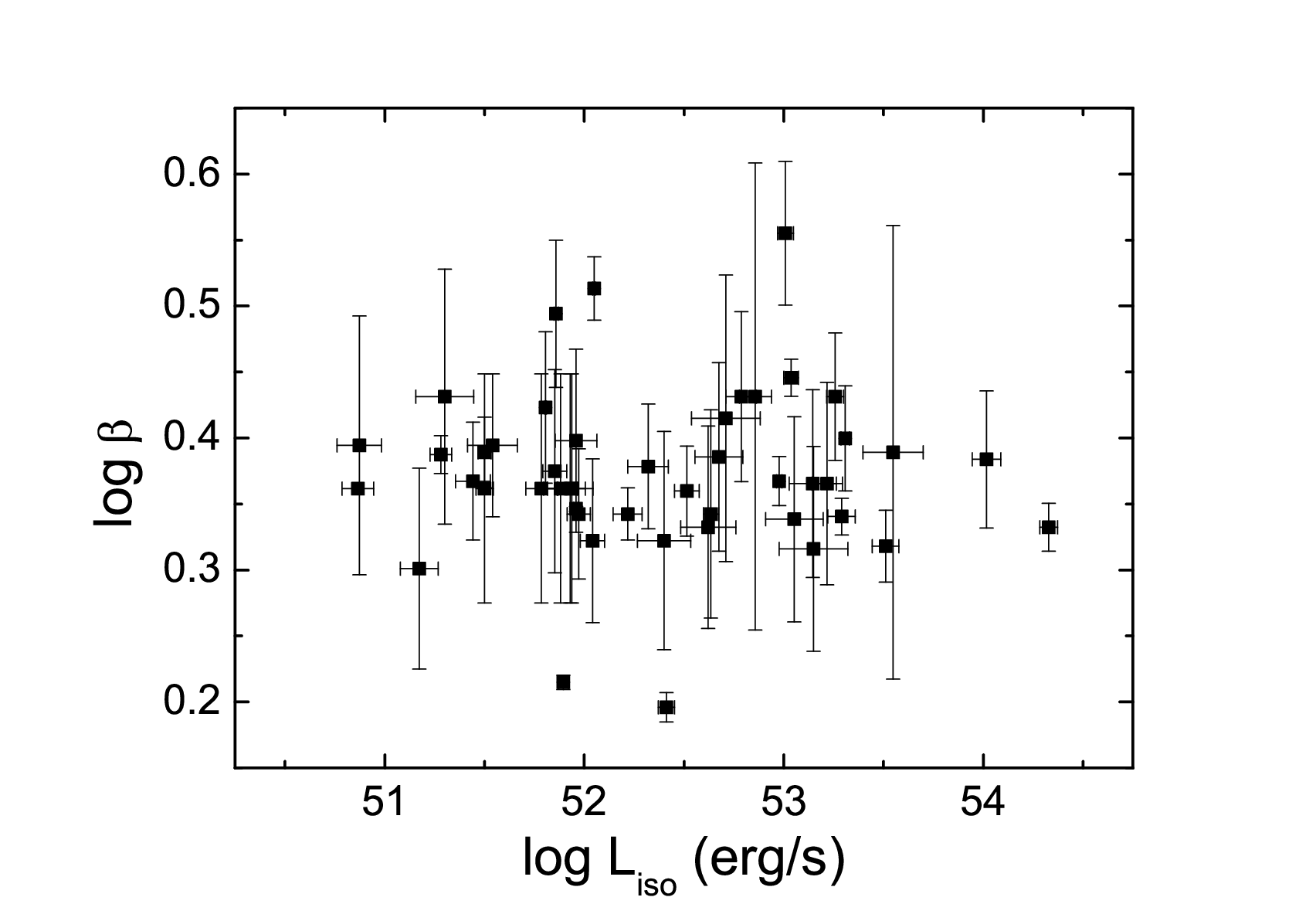}
   \caption{Left panel: $\alpha$ vs $L_{\rm iso}$ for the GRBs in our sample. The solid line
            corresponds to the best fit. Right panel: the corresponding $\beta$ vs
            $L_{\rm iso}$ diagram. }
   \label{Fig:plot7}
\end{figure}

\begin{figure}
   \plottwo{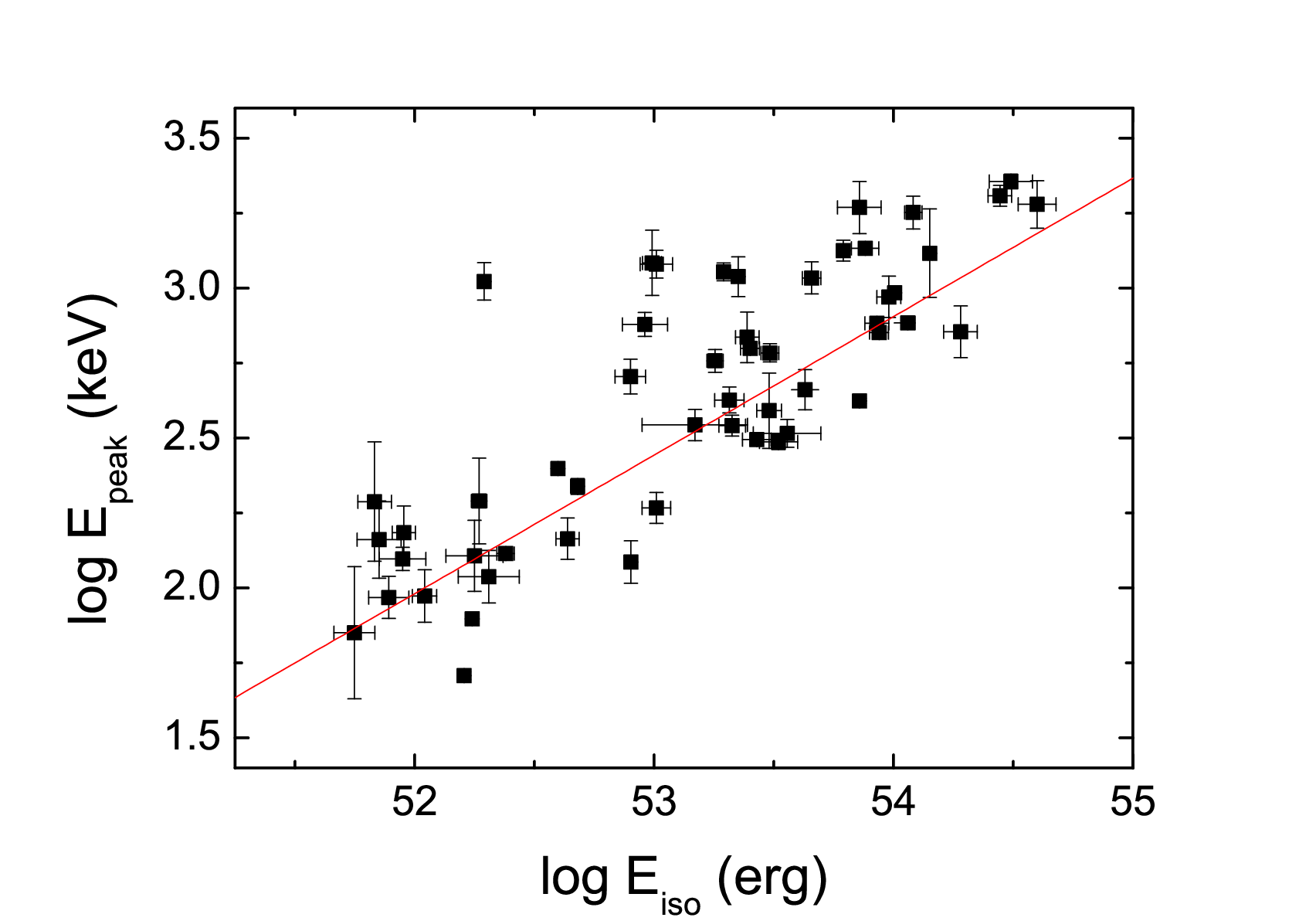}{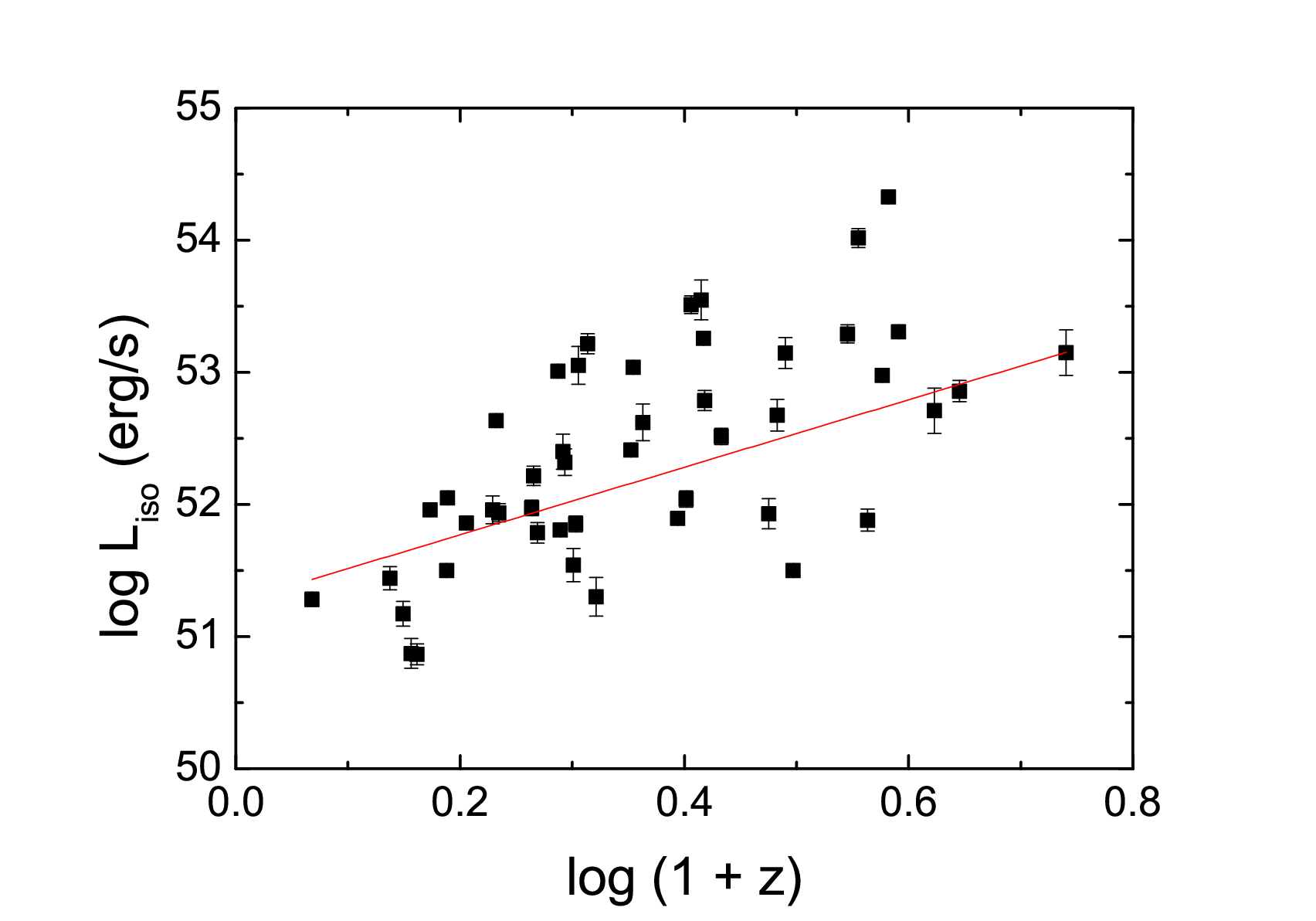}
   \caption{Left panel: the Amati relation ($E_{\rm peak}$ vs $E_{\rm iso}$) for the GRBs in our sample.
            The solid line corresponds to the best fit. Right panel: $L_{\rm iso}$ vs $z$ for the
            same sample. The solid line is our best fit. }
   \label{Fig:plot8}
\end{figure}

\begin{figure}
   \begin{center}
   \plotone{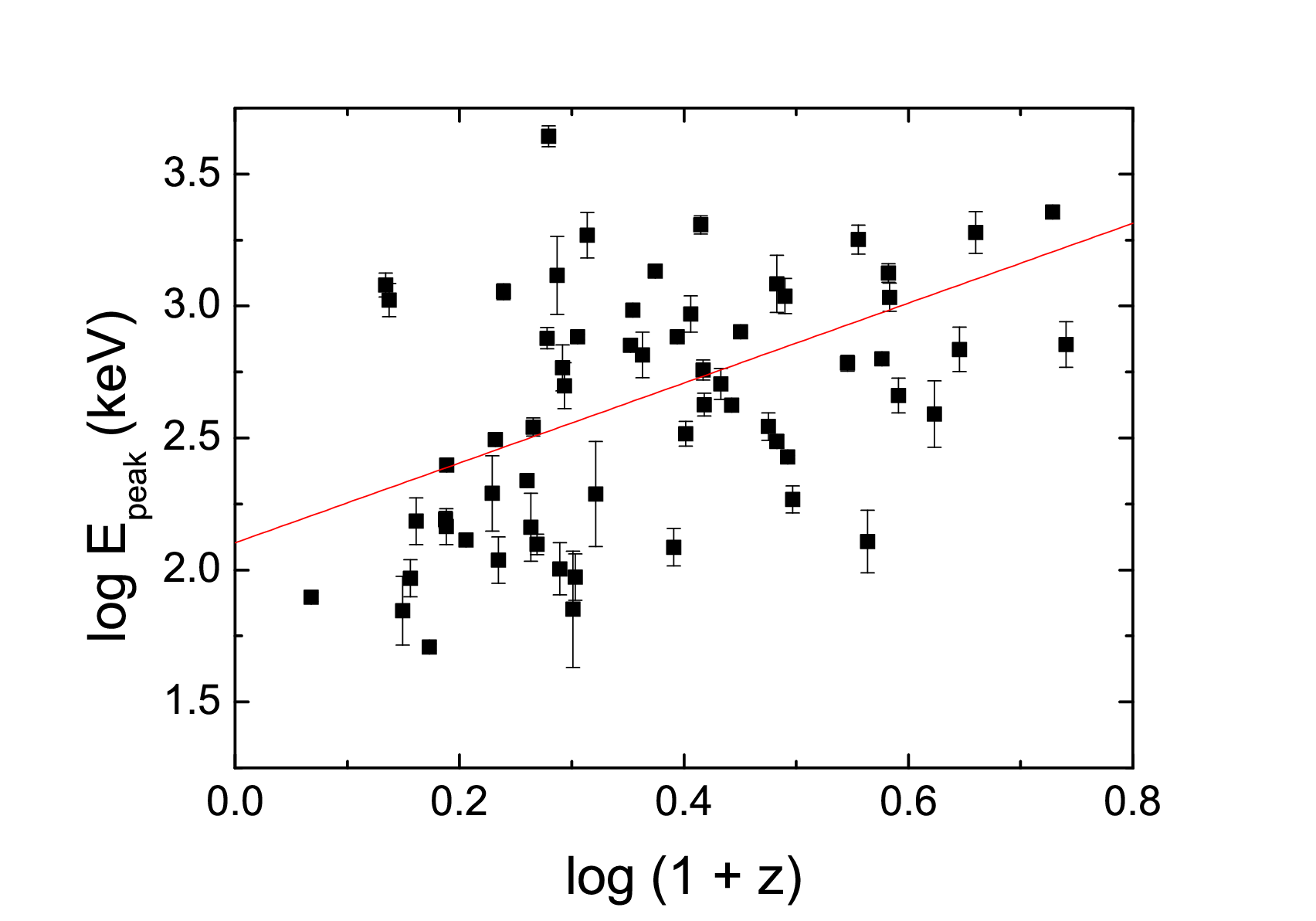}
   \caption{$E_{\rm peak}$ (rest frame) vs $z$ for the GRBs in our sample. The solid line corresponds to
            the best fit result.}
   \label{Fig:plot9}
   \end{center}
\end{figure}

\begin{figure}
   \begin{center}
   \includegraphics[width=10.0cm]{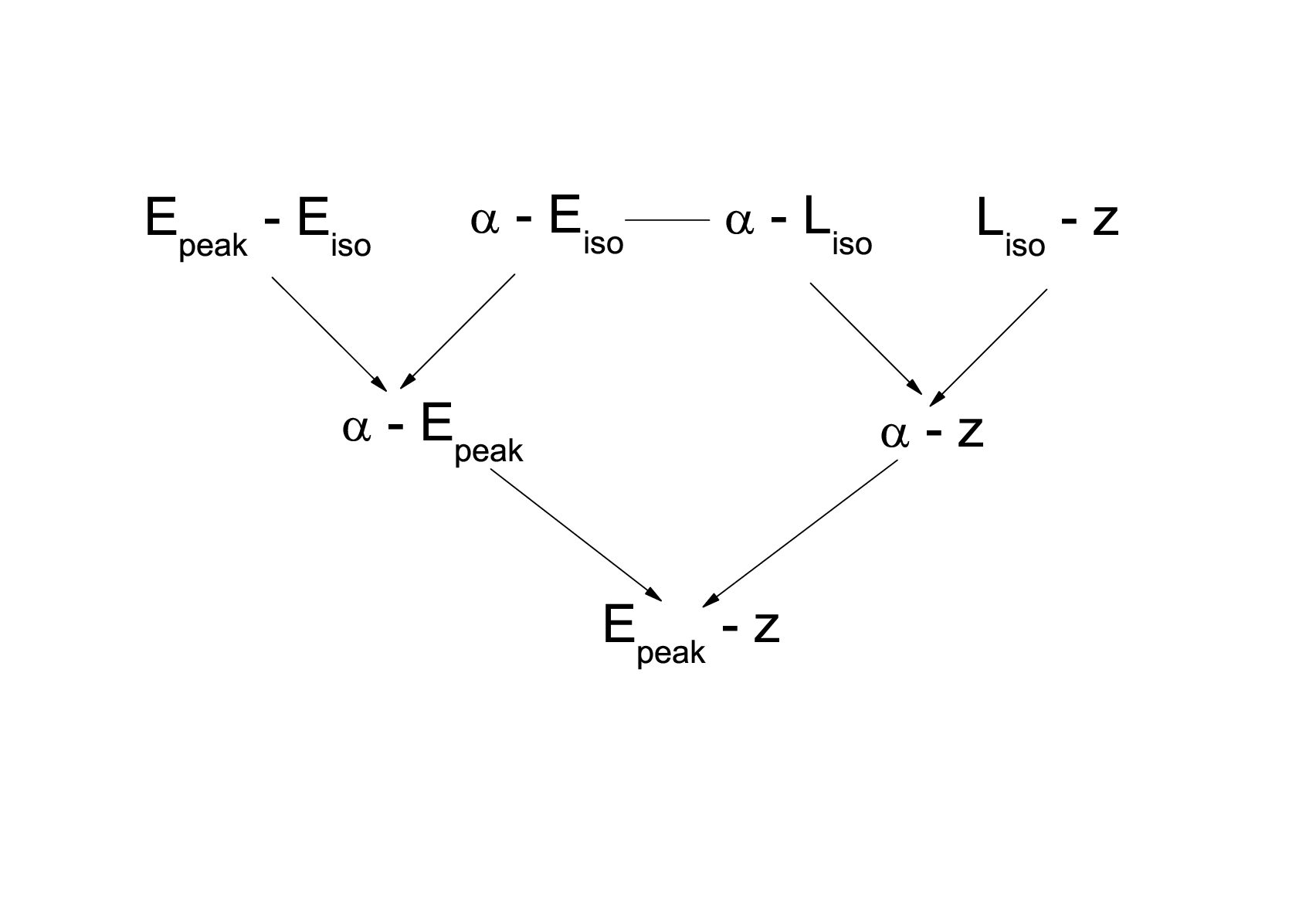}
   \caption{A schematic illustration of the relationship of the correlations studied in our work.}
   \label{Fig:plot10}
   \end{center}
\end{figure}

\begin{figure}
   \vspace{0.5cm}
   \begin{center}
   \plotone{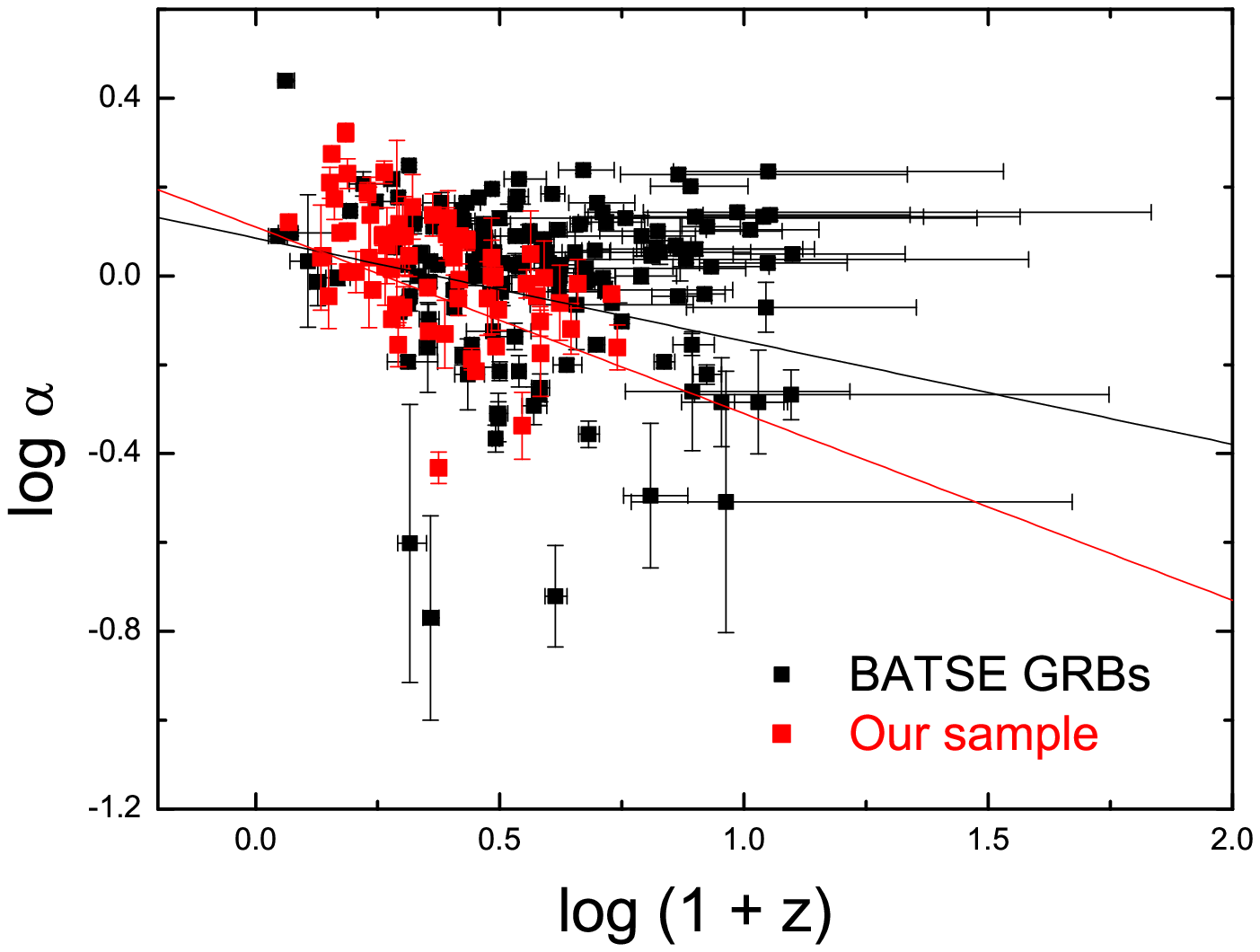}
   \caption{A comparison of our sample (red dots) and GRBs from the BATSE catalog
            (black dots) on the $\log \alpha - \log (1 + z)$ plane. The pseudo redshifts of BATSE GRBs
            are derived from the Yonetoku relation (Yonetoku et al.~\citeyear{yone04}).
            The solid lines are the fitting results of our sample (red) and BATSE GRBs (black). }
   \label{Fig:plot11}
   \end{center}
\end{figure}

\clearpage

\begin{deluxetable}{cccccccccc}
\tablecaption{List of GRBs with known redshifts and Band function spectral parameters that define our sample.}
\tabletypesize{\scriptsize}
\tabcolsep 2.5pt
\tablehead{
\colhead{Name}       & \colhead{$z$}    &
\colhead{$\alpha$}         & \colhead{$\beta$}  &
\colhead{Range}          & \colhead{$E_{\rm peak}$\tablenotemark{a}}  &
\colhead{$E_{\rm iso}$\tablenotemark{b}}  & \colhead{$L_{\rm iso}$\tablenotemark{c}}  &
\colhead{Mission\tablenotemark{d}} & \colhead{Ref.}\\
 &   &          &         & \colhead{(keV)} & \colhead{(keV)} & \colhead{($10^{\rm 52}erg$)} & \colhead{($10^{\rm 51}erg/s$)} &  &
}
\startdata
970228  &   0.695   &   1.54    $\pm$   0.08    &   2.5 $\pm$   0.4 &   2-700   &   195 $\pm$   64  &   1.86    $\pm$    0.14   &   9.1 $\pm$   2.18    &   SAX/WFC &   1,2 \\
970508  &   0.835   &   1.71    $\pm$   0.1 &   2.2 $\pm$   0.25    &   2-700   &   145 $\pm$   43  &   0.71    $\pm$    0.15   &   9.4 $\pm$   1.25    &   SAX/WFC &   1,2 \\
970828  &   0.958   &   0.7 $\pm$   0.08    &   2.1 $\pm$   0.4 &   30-10000    &   583 $\pm$   117 &           &   25.1     $\pm$  7.7 &   RXTE/ASM    &   2   \\
971214  &   3.42    &   0.76    $\pm$   0.1 &   2.7 $\pm$   1.1 &   2-700   &   685 $\pm$   133 &   24.5    $\pm$   2.8  &  72.1    $\pm$   13.3    &   SAX/WFC &   1,2 \\
980326  &   1   &   1.23    $\pm$   0.21    &   2.48    $\pm$   0.31    &   2-700   &   71  $\pm$   36  &   0.56     $\pm$  0.11    &   3.47    $\pm$   1   &   SAX/WFC &   1,2 \\
980613  &   1.096   &   1.43    $\pm$   0.24    &   2.7 $\pm$   0.6 &   2-700   &   194 $\pm$   89  &   0.68    $\pm$    0.11   &   2   $\pm$   0.67    &   SAX/WFC &   1,2 \\
980703  &   0.966   &   1.31    $\pm$   0.14    &   2.39    $\pm$   0.26    &   50-300  &   499 $\pm$   100 &            &  20.9    $\pm$   4.86    &   RXTE/ASM    &   2   \\
990123  &   1.6 &   0.89    $\pm$   0.08    &   2.45    $\pm$   0.97    &   2-700   &   2030    $\pm$   161 &   278.3    $\pm$  31.5    &   353 $\pm$   123 &   SAX/WFC &   1,2 \\
990506  &   1.307   &   1.37    $\pm$   0.15    &   2.15    $\pm$   0.38    &   50-300  &   653 $\pm$   130 &            &  41.8    $\pm$   13.3    &   BATSE   &   2   \\
990510  &   1.619   &   1.23    $\pm$   0.05    &   2.7 $\pm$   0.4 &   2-700   &   423 $\pm$   42  &   20.6    $\pm$    2.9    &   61.2    $\pm$   10.7    &   SAX/WFC &   1,2 \\
990705  &   0.843   &   1.05    $\pm$   0.21    &   2.2 $\pm$   0.1 &   2-700   &   348 $\pm$   28  &   21.2    $\pm$    2.7    &   16.5    $\pm$   2.77    &   SAX/WFC &   1,2 \\
990712  &   0.433   &   1.88    $\pm$   0.07    &   2.48    $\pm$   0.56    &   2-700   &   93  $\pm$   15  &   0.78     $\pm$  0.15    &   0.746   $\pm$   0.191   &   SAX/WFC &   1,2 \\
991208  &   0.706   &   1.1 $\pm$   0.4 &   2.2 $\pm$   0.4 &   20-2000 &   312.3   $\pm$   5.1 &   26.9    $\pm$   3.7  &  43.2    $\pm$   3.8 &   K/W  &   2,3  \\
991216  &   1.02    &   1.23    $\pm$   0.25    &   2.18    $\pm$   0.39    &   20-2000 &   763.6   $\pm$   20.2    &    85.1   $\pm$   9.8 &   113 $\pm$   37.5    &   BATSE   &   2,3  \\
000131  &   4.5 &   0.69    $\pm$   0.08    &   2.07    $\pm$   0.37    &   50-300  &   714 $\pm$   142 &   190.5    $\pm$  30.7    &   141 $\pm$   55.9    &   BATSE   &   2,3 \\
000214  &   0.42    &   1.62    $\pm$   0.13    &   2.1 $\pm$   0   &   40-700  &               &   0.8 $\pm$   0.026    &          &   SAX/WFC &   4   \\
000301C &   2.04    &   1   $\pm$   0.3 &   2.3 $\pm$   0.3 &   20-2000 &   1213    $\pm$   303 &   9.8 $\pm$   0.9 &            &  SAX/WFC &   3   \\
000911  &   1.06    &   1.11    $\pm$   0.12    &   2.32    $\pm$   0.41    &   15-8000 &   1856    $\pm$   371 &   72   $\pm$  15  &   165 $\pm$   28.9    &   IPN &   2   \\
000926  &   2.04    &   1.1 $\pm$   0.1 &   2.43    $\pm$   0.4 &   20-2000 &   306.9   $\pm$   18.2    &   33.1     $\pm$  6.1 &   47.3    $\pm$   13  &   K/W  &   2,3 \\
010222  &   1.477   &   1.35    $\pm$   0.19    &   1.64    $\pm$   0.02    &   20-2000 &   765.4   $\pm$   29.7    &    114.8  $\pm$   7.9 &   7.87    $\pm$   0.45    &   SAX/WFC &   2,3 \\
010921  &   0.45    &   1.49    $\pm$   0.16    &   2.3 $\pm$   0   &   30-700  &   153 $\pm$   31  &   0.9 $\pm$   0.1  &  0.733   $\pm$   0.133   &   HETE-2  &   2,4 \\
011121  &   0.362   &   1.1 $\pm$   0.3 &   2.1 $\pm$   0.2 &   20-2000 &   1201    $\pm$   126.7   &   10.23   $\pm$    1.6    &           &   SAX/WFC &   3   \\
011211  &   2.14    &   0.84    $\pm$   0.09    &   2.3 $\pm$   0.46    &   40-700  &   185 $\pm$   22  &   10.23    $\pm$  1.4 &   3.17    $\pm$   0.32    &   SAX/WFC &   2,3 \\
020124  &   3.198   &   0.87    $\pm$   0.17    &   2.6 $\pm$   0.65    &   2-400   &   390 $\pm$   113 &   30.2     $\pm$  3.6 &   51.2    $\pm$   20.3    &   HETE-2  &   2,4 \\
020813  &   1.25    &   0.94    $\pm$   0.03    &   1.57    $\pm$   0.04    &   20-2000 &   710 $\pm$   33.8    &   87   $\pm$  8   &   25.8    $\pm$   2.4 &   HETE-2  &   2,3 \\
020819B &   0.41    &   0.9 $\pm$   0.15    &   2   $\pm$   0.35    &   25-100  &   70  $\pm$   21  &           &   1.49     $\pm$  0.323   &   HETE-2  &   2   \\
021211  &   1.01    &   0.85    $\pm$   0.09    &   2.37    $\pm$   0.42    &   2-400   &   94  $\pm$   19  &   1.1  $\pm$  0.13    &   7.13    $\pm$   0.99    &   HETE-2  &   2,4 \\
030226  &   1.987   &   0.89    $\pm$   0.17    &   2.3 $\pm$   0.46    &   20-2000 &   349.5   $\pm$   41.8    &   14.8     $\pm$  7.5 &   8.52    $\pm$   2.23    &   HETE-2  &   2,3 \\
030328  &   1.52    &   1.14    $\pm$   0.03    &   2.1 $\pm$   0.3 &   2-400   &   328 $\pm$   35  &   36  $\pm$   11.7     &  11  $\pm$   1.55    &   HETE-2  &   2,3 \\
030329  &   0.169   &   1.32    $\pm$   0.02    &   2.44    $\pm$   0.08    &   2-400   &   79  $\pm$   3   &   1.74     $\pm$  0.08    &   1.91    $\pm$   0.237   &   HETE-2  &   2,3 \\
030429  &   2.66    &   1.12    $\pm$   0.25    &   2.3 $\pm$   0.46    &   2-400   &   128 $\pm$   35  &   1.78     $\pm$  0.49    &   7.6 $\pm$   1.47    &   HETE-2  &   2,3 \\
040924  &   0.858   &   1.17    $\pm$   0.23    &   2.3 $\pm$   0.46    &   20-500  &   125 $\pm$   11  &   0.89     $\pm$  0.2 &   6.1 $\pm$   1.1 &   HETE-2  &   2,3 \\
041006  &   0.716   &   1.37    $\pm$   0.27    &   2.3 $\pm$   0.46    &   30-400  &   109 $\pm$   22  &   2.04     $\pm$  0.6 &   8.65    $\pm$   1.36    &   HETE-2  &   2,3 \\
050401  &   2.9 &   0.99    $\pm$   0.19    &   2.51    $\pm$   0.23    &   20-2000 &   458.2   $\pm$   70.2    &   42.7     $\pm$  2.9 &   203 $\pm$   10  &   K/W  &   2,3 \\
050525A &   0.606   &   1.02    $\pm$   0.11    &   3.12    $\pm$   0.4 &   15-10000    &   130.1   $\pm$   4.8 &   2.4  $\pm$  0.2 &   7.23    $\pm$   0.18    &   K/W  &   3,5 \\
050603  &   2.821   &   0.79    $\pm$   0.06    &   2.15    $\pm$   0.09    &   20-3000 &   1334    $\pm$   107 &   61.7     $\pm$  1.4 &   2130    $\pm$   220 &   K/W  &   2,3 \\
061007  &   1.261   &   0.75    $\pm$   0.02    &   2.79    $\pm$   0.09    &   20-10000    &   965 $\pm$   27  &   101  $\pm$  1.4 &   109 $\pm$   9.1 &   K/W  &   5   \\
061222A &   2.09    &   1   $\pm$   0.05    &   2.32    $\pm$   0.38    &   20-10000    &   1091    $\pm$   167 &   22.5     $\pm$  0.94    &   140 $\pm$   38  &   K/W  &   5   \\
070125  &   1.547   &   1.1 $\pm$   0.1 &   2.08    $\pm$   0.13    &   20-10000    &   934 $\pm$   148 &   95.5     $\pm$  11  &   324 $\pm$   50  &   K/W  &   2,3 \\
070328  &   0.372   &   1.11    $\pm$   0.04    &   2.33    $\pm$   0.24    &   20-10000    &   1052    $\pm$   152 &    1.95   $\pm$   0.07    &   2.77    $\pm$   0.56    &   K/W  &   5   \\
071010B &   0.947   &   1.25    $\pm$   0.6 &   2.65    $\pm$   0.35    &   20-1000 &   101 $\pm$   23  &                &  6.4 $\pm$   0.053   &   K/W  &   2   \\
080319B &   0.937   &   0.86    $\pm$   0.01    &   3.59    $\pm$   0.45    &   20-7000 &   1307    $\pm$   443 &   142  $\pm$  3   &   102 $\pm$   9.4 &   K/W  &   5   \\
080721  &   2.591   &   0.96    $\pm$   0.07    &   2.42    $\pm$   0.29    &   20-7000 &   1785    $\pm$   223 &   121  $\pm$  10  &   1038    $\pm$   172 &   K/W  &   5   \\
080916C &   4.35    &   0.91    $\pm$   0.02    &   2.08    $\pm$   0.06    &   10-1000000  &   2268    $\pm$   128 &    309    $\pm$   64  &           &   Fermi   &   3   \\
081007  &   0.53    &   2.1 $\pm$   0.1 &   10  $\pm$   10  &   25-900  &               &               &            &   Fermi  &   6   \\
081121  &   2.512   &   0.46    $\pm$   0.08    &   2.19    $\pm$   0.07    &   8-35000 &   608 $\pm$   42  &   30.5     $\pm$2.6   &   195 $\pm$   31  &   Fermi   &   5   \\
081222  &   2.77    &   0.9 $\pm$   0.03    &   2.33    $\pm$   0.1 &   8-35000 &   630 $\pm$   31  &   25.2    $\pm$    2.3    &   94.9    $\pm$   3.1 &   Fermi   &   5   \\
090323  &   3.57    &   0.96    $\pm$   0.12    &   2.09    $\pm$   0.19    &   20-10000    &   1901    $\pm$   347 &    398    $\pm$   73  &               &   K/W  &   3   \\
090328  &   0.735   &   0.93    $\pm$   0.02    &   2.2 $\pm$   0.1 &   8-1000  &   1133    $\pm$   78  &   19.5     $\pm$0.9   &               &   Fermi   &   3   \\
090424  &   0.544   &   1.02    $\pm$   0.01    &   3.26    $\pm$   0.18    &   8-35000 &   250 $\pm$   3.4 &   3.97     $\pm$0.08  &   11.2    $\pm$   0.17    &   Fermi   &   5   \\
090510  &   0.903   &   0.8 $\pm$   0.03    &   2.6 $\pm$   0.3 &   8-40000 &   4400    $\pm$   400 &               &                &  Fermi   &   6   \\
090618  &   0.54    &   1.26    $\pm$   0.04    &   2.5 $\pm$   0.25    &   8-1000  &   155.5   $\pm$   11  &                &              &   Fermi   &   6   \\
090902B &   1.822   &   0.61    $\pm$   0.01    &   3.87    $\pm$   0.16    &   8-1000  &   798 $\pm$   7   &                &              &   Fermi   &   6   \\
090926A &   2.1062& 0.693   $\pm$   0.01    &   2.34    $\pm$   0.011   &   8-1000  &   268 $\pm$   4   &                 &             &   Fermi   &   6   \\
091003  &   0.8969& 1.04    $\pm$0.05   &   3.58    $\pm$1.21   &   8-35000 &   755.7   $\pm$   70  &   9.16    $\pm$    2  &               &   Fermi&  7   \\
091020  &   1.71    &   1.2 $\pm$   0.06    &   2.29    $\pm$   0.18    &   8-35000 &   507 $\pm$   68  &   7.96     $\pm$1.16  &   32.7    $\pm$   4.6 &   Fermi   &   5   \\
091127  &   0.49    &   1.25    $\pm$   0.05    &   2.22    $\pm$   0.01    &   8-35000 &   51  $\pm$   1.5 &   1.61     $\pm$0.03  &   9.08    $\pm$   0.22    &   Fermi   &   5   \\
100414A &   1.368&  0.37    $\pm$   0.03    &   3.74    $\pm$   0.95    &   8-35000 &   1356.4$\pm$38.4 &   76.2     $\pm$  10  &               &   Fermi&  7   \\
100621A &   0.542   &   1.7 $\pm$   0.13    &   2.45    $\pm$   0.15    &   20-2000 &   146 $\pm$   23  &   4.35     $\pm$0.48  &   3.17    $\pm$   0.24    &   K/W  &   5   \\
100814A &   1.44 &  0.74    $\pm$   0.13    &   2.73    $\pm$   0.69    &   8-35000 &               &               &                &  Fermi&  7    \\
110213A &   1.46    &   1.24    $\pm$   0.17    &   2.08    $\pm$   0.05    &   8-35000 &   122  $\pm$  20  &   8    $\pm$  0.4   &             &   Fermi&  7    \\
110422A &   1.77    &   0.65    $\pm$   0.036&  2.96    $\pm$   0.12    &   20-5000 &   421 $\pm$   8   &   72  $\pm$    0.3    &               &   K/W& 7    \\
110503A &   1.613   &   0.98    $\pm$   0.08    &   2.7 $\pm$   0.3 &   20-5000 &   572 $\pm$   50  &   18  $\pm$   1.4  &  181 $\pm$   18  &   K/W  &   5   \\
110715A &   0.82    &   1.23    $\pm$   0.06    &   2.7 $\pm$   0.3 &   20-5000 &   218 $\pm$13 &   4.8 $\pm$   0.2 &                &  K/W& 7    \\
110731A &   2.83    &   0.67    $\pm$   0.15    &   2.64    $\pm$   0.42    &   8-35000 &   1080    $\pm$   133 &   45.5     $\pm$  4   &               &   Fermi&      7    \\
\enddata
  \tablenotetext{a}{$E_{\rm peak}$ is the peak energy of $\nu F_{\nu}$ spectrum in the GRB rest frame.}
  \tablenotetext{b}{The integrating range of $E_{\rm iso}$ is 1 -- 10$^4$ keV in the GRB rest frame.}
  \tablenotetext{c}{The integrating range of $L_{\rm iso}$ is 1 -- 10$^4$ keV in the GRB rest frame.}
  \tablenotetext{d}{The name of the mission from which the spectral parameters have been
                    derived: SAX/WFC = BeppoSAX, K/W = Konus/Wind, IPN = InterPlanetary Network.}
\tablerefs{
(1) Amati et al.~\citeyearpar{amati02}; (2) Nava et al.~\citeyearpar{nava08}; (3) Kann et al.~\citeyearpar{kann10}; (4) Ghirlanda et al.~\citeyearpar{ghir04}; (5) Nava et al.~\citeyearpar{nava12}; (6) Nava et al.~\citeyearpar{nava11b}; (7) Tsutsui et al.~\citeyearpar{tsut12}.}
\end{deluxetable}

\end{document}